\documentclass[aps,prb,twocolumn,showpacs,amsmath,amssymb,floatfix]{revtex4}
\usepackage{color}
\usepackage{epsfig} \usepackage{bm}

\IfFileExists{srcltx.sty}{\usepackage[active]{srcltx}}

\begin{document}

\title{Theory of fractional quantum Hall interferometers}

\author{Ivan P. Levkivskyi$^{1,2}$, J\"{u}rg Fr\"{o}hlich$^3$ and Eugene V. Sukhorukov$^1$}
\affiliation{$^1$D\'epartement de Physique Th\'eorique, Universit\'e de
Gen\`eve, CH-1211 Gen\`eve 4, Switzerland}
\affiliation{$^2$Physics Department, Kyiv National University, 03022 Kyiv, Ukraine}
\affiliation{$^3$Institute of Theoretical Physics, ETH H\"{o}nggerberg, CH-8093 Zurich, Switzerland}

\begin{abstract}
Interference of fractionally charged quasi-particles is expected to lead to Aharonov-Bohm oscillations
with periods larger than the flux quantum. However, according to the Byers-Yang theorem, observables of an
electronic system are invariant under an adiabatic insertion of a quantum of singular flux. We resolve this seeming
paradox by considering a microscopic model of electronic interferometers made from a quantum Hall liquid at
filling factor $1/m$ with the shape of a Corbino disk. In such interferometers, the quantum Hall edge states are utilized in place of optical
beams, the quantum point contacts play the role of beam splitters connecting different edge channels, and Ohmic contacts
represent a source and drain of quasi-particle currents. Depending on the position of Ohmic contacts one
distinguishes interferometers of Fabry-P\'{e}rot (FP)  and Mach-Zehnder (MZ) type. An approximate ground state of such
interferometers is described by a Laughlin type wave function, and low-energy excitations are incompressible deformations
of this state. We construct a low-energy effective theory by restricting the microscopic Hamiltonian of electrons to the space of
incompressible deformations and show that the theory of the quantum Hall edge so obtained is a generalization of a chiral
conformal field theory. In our theory, a quasi-particle tunneling operator is found to be a single-valued function of tunneling
point coordinates, and its phase depends on the topology determined by the positions of Ohmic contacts. We describe strong
coupling of the edge states to Ohmic contacts and the resulting quasi-particle current through the interferometer with
the help of a master equation. We find that 
the coherent contribution to the average quasi-particle current through MZ
interferometers does not vanish after summation over quasi-particle degrees of freedom. However, it acquires oscillations with the electronic period, in agreement with the Byers-Yang theorem.
Importantly, our theory does not rely on any ad-hoc constructions, such as Klein factors,
etc. When the magnetic flux through an FP interferometer is varied with a modulation gate, current oscillations have the
quasi-particle periodicity, thus allowing for spectroscopy of quantum Hall edge states. 
\end{abstract}


\maketitle

\section{Introduction.}

Since its discovery, in 1980, the quantum Hall (QH) effect \cite{Klitzing} has been a very rich source
of interesting problems related to topological and correlation effects in condensed matter systems.
The QH effect is observed in two-dimensional electron gases\cite{Datta} (2DEG) exposed to a strong homogeneous magnetic
field perpendicular to the plane of the gas. At appropriate electron densities, the 2DEG forms an
incompressible liquid. The QH effect manifests itself in the precise and universal quantization
of the Hall conductance. This behavior originates from an interplay between the Landau quantization of
the orbital motion of electrons \cite{Landau} and interaction effects, which leads to the formation of a
bulk energy gap. As a consequence, an incompressible state is formed in the bulk of the 2DEG. At the edge
of a 2DEG exhibiting the QH effect there exist, however, gapless chiral modes that are
the quantum analogue of classical skipping orbits.\cite{QHE} In the presence of strong Coulomb
interactions, these modes can be viewed as collective edge plasmon modes. Remarkably,
it has been predicted\cite{Wen,Frol} that, at fractional fillings of the Landau levels, besides the collective
modes the edge states of the 2DEG also describe quasi-particles with {\em fractional charges}
and fractional statistics. \cite{Laugh} For instance, in QH liquids with filling factor $\nu = 1/m$,
where $m$ is an odd integer, Laughlin quasi-particle excitations have
an electric charge $e^* = e/m$, where $e$ is the elementary electric charge, and fractional statistics. Such excitations
can be scattered between opposite edges at narrow constrictions forming quantum point contacts (QPC),
thus contributing to a backscattering current.

The fractional charge of Laughlin quasi-particles has been confirmed experimentally in measurements
of the shot noise of weak backscattering currents.  \cite{fract-charge-meas} The quasi-particle charge
in these experiments is inferred from the Fano factor of noise, which is the ratio of the noise
power to the average backscattering current. Although, at present, there is a consensus on the
interpretation of the experimental results,  this type of measurement does
not, in general, represent a direct test of the fractional charge of quasi-particles.  \cite{footnote0}
Indeed, the Fano factor of noise
is not universal and may be reduced or enhanced for various reasons.  \cite{Blanter}
For instance, the so called ``charge fractionalization'' in nonchiral one-dimensional
systems \cite{Ines,Yacoby} is a property of collective modes that has nothing to do with
the existence of fractionally charged quasi-particles. Nevertheless, this phenomenon reduces the Fano
factor of noise at relatively high frequencies.  \cite{footnote1}

\begin{center}\begin{figure}[h]\begin{center}
\epsfxsize=7cm
\epsfbox{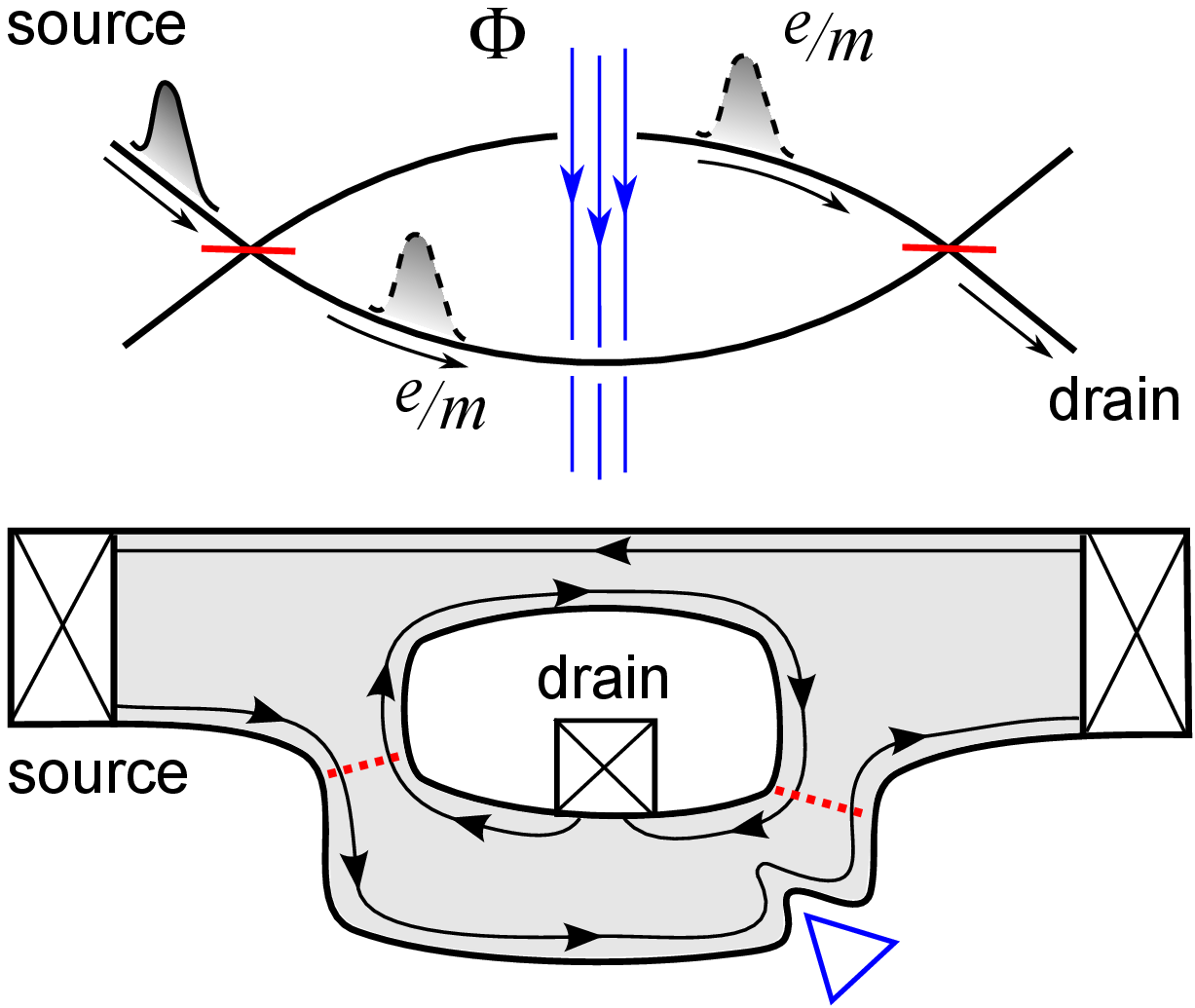}
\caption{An electronic analogue of an optical Mach-Zehnder (MZ) interferometer is shown
schematically. {\em Upper panel:} According to a gedanken formulation of
the Aharonov-Bohm  effect, quasi-particles with fractional charge $e/m$
propagate from a source to a drain via two beam splitters, and enclose a singular
magnetic flux $\Phi$. {\em Lower panel:} Schematic sketch of a typical experimental
realization of the MZ interferometer in a quantum Hall system.
 \cite{MZ-experiment1,mz1, MZ-experiment3, MZ-experiment4, MZ-experiment2}
Chiral edge states, shown by arrows, play the role of optical beams. They are split at
two quantum point contacts indicated by dashed lines. The source and drain are Ohmic
contacts. The quantum Hall liquid is confined to a region with the topology of a Corbino disk (indicated by gray shading).
The magnetic flux through the interferometer is typically changed by a modulation gate, shown as a blue triangle. However,
it is possible, at least in principle,
to insert a singular magnetic flux through the hole in the Corbino disk.}
\vspace{-5mm}
\label{frol-simpl}
\end{center}\end{figure}\end{center}

A direct measurement of the quasi-particle charge should rely on the transformation properties
of the quasi-particle operators under electromagnetic gauge transformations and on the quantum 
nature of quasi-particles. The most appealing approach is to make use of the 
Aharonov-Bohm (AB) effect,  \cite{AB} which relies on the fact that the
interference of quasi-particles is affected by a magnetic flux. Following a commonly used
formulation of this effect, we consider a gedanken interference experiment shown in the
upper panel of Fig.\ \ref{frol-simpl}. Quasi-particles of charge $e/m$ traverse two
beam splitters and follow paths that enclose a {\em singular} magnetic flux $\Phi$. The relative
phase between the two amplitudes, for the upper and lower path, is shifted by an amount of
$2\pi\Phi/m\Phi_0$, which leads to AB oscillations in the current from the source to the drain as a function
of the flux $\Phi$ with quasi-particle period $m\Phi_0$. Then the quasi-particles can be
detected by differentiating their contribution to the AB effect from electron oscillations with period $\Phi_0$.
Such gedanken experiment can be realized in electronic QH interferometers, 
studied extensively in recent experimental\cite{MZ-experiment1,mz1, MZ-experiment3, MZ-experiment4, MZ-experiment2,
FP1,FP2,FP3,FP4} and theoretical\cite{MZ-theory1, MZ-theory2, MZ-theory3,MZ-theory4, our,our-frac, MZ-theory5, Rosenow} works, in particular,  in a Mach-Zehnder (MZ) interferometer shown schematically 
in the lower panel of Fig.\ \ref{frol-simpl}. 
The gedanken formulation of the AB effect however leads to a paradox:
In any electronic system, including fractional QH systems, AB oscillations
should have an electronic period $\Phi_0$, according to the Byers-Yang theorem.  \cite{BY-theorem}
This is so, because, after adiabatic insertion of a flux quantum through a hole in the sample, 
an electronic system relaxes  to its initial state, since the flux quantum can be removed by a 
single-valued gauge transformation.

\begin{center}\begin{figure}[hbt]
\epsfxsize=7cm
\begin{center}
\epsfbox{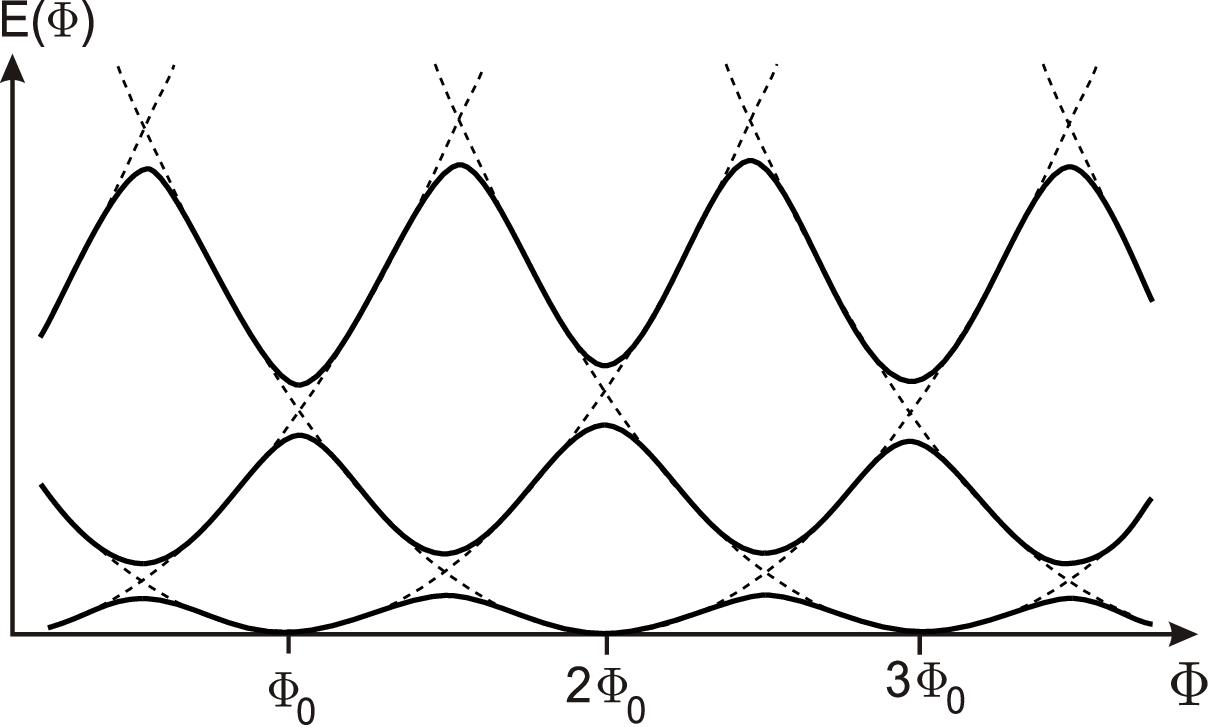}
\caption{The energy spectrum of a QH fluid at $\nu=1/3$ in a Corbino disk
is schematically shown. The dashed lines show the Coulomb charging energy
of an isolated QH fluid, as a function of the magnetic flux $\Phi$ threading
the Corbino disk. Different
branches correspond to different numbers of quasi-particles at the edges.
It is suggested in Ref.\  [\onlinecite{Thouless}] 
that in the presence of inter-edge quasi-particle tunneling, and for weak coupling
to metallic reservoirs, an energy gap opens at the degeneracy points, where
different branches intersect. The authors then argue that 
if the flux varies adiabatically, the QH fluid
follows the ground state, so that the energy is a periodic function of the flux
with the electronic period $\Phi_0$.}
\vspace{-2mm}
\label{magn}\end{center}
\end{figure}\end{center}

There have been several theoretical attempts to resolve this paradox.
\cite{Kane,Feldman,Averin,Thouless,Feldman2}
In early work,  \cite{Thouless} Thouless and Gefen have considered the energy spectrum
of a QH liquid confined to a Corbino disk and weakly coupled to Ohmic contacts, see Fig.\ \ref{magn}. They have
found that, as a result of weak quasi-particle tunneling between the inner and the outer edge
of the Corbino disk, the energy spectrum, and consequently, ``any truly thermodynamic quantity''
is a periodic function of the magnetic flux with the electronic period $\Phi_0$. Although
Thouless and Gefen have made an important first step towards understanding the AB effect in
QH interferometers, their analysis
is rather qualitative, and some of their statements concerning the tunneling
rates and currents are not firmly justified.  \cite{footnote2}
The results of their work are somewhat difficult to interpret at the level of effective theories. But, more
importantly, they cannot easily be generalized to the situation where the magnetic flux is varied
with the help of a gate voltage. This situation has stimulated further
interest in the fractional AB effect.

More recently, a number of authors (see Refs.\  [\onlinecite{Kane,Feldman,Feldman2}]) 
have proposed that the correct
description of the MZ interferometers should take into account additional quantum
numbers labeling a QH state. In the thermodynamic limit, 
averaging the quasi-particle current over such quantum
numbers is claimed to restore the electronic AB periodicity. These quantum numbers are usually introduced
ad hoc, with minimal justification. They are
represented either in terms of so called Klein factors in the tunneling operators,  \cite{Safi} or
as additional phase shifts induced by quasi-particles localized at the inner edge of the Corbino disk.
 \cite{Feldman}
We are aware of only one attempt to justify theoretically the introduction of Klein factors
in MZ interferometers: In their recent work,  \cite{Averin} Ponomarenko and Averin propose
a resummation of electron tunneling processes between the inner and outer edge and claim to unravel a duality
to weak quasi-particle tunneling, where the Klein factors arise naturally. However, their
analysis is carried out entirely at the level of an effective theory, where weak  quasi-particle
and weak electron tunneling are the two fixed points of a renormalization group flow.
There is no guarantee that microscopic considerations yield the same result.

\begin{center}\begin{figure}[htb]\begin{center}
\epsfxsize=5cm
\epsfbox{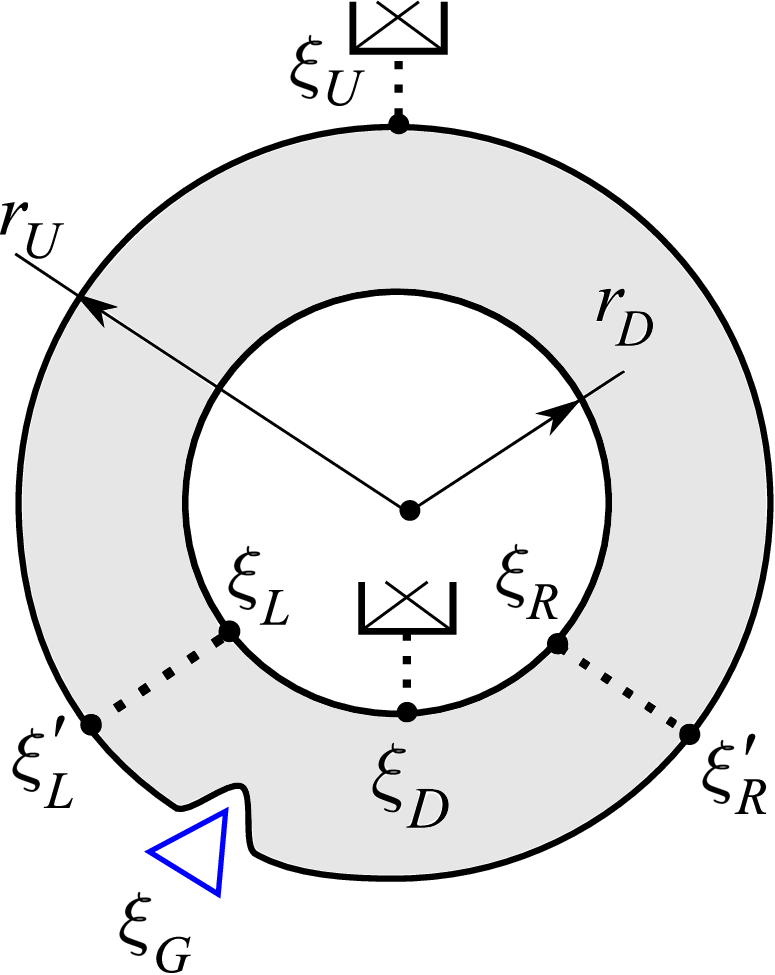}
\caption{Illustration of our model of a QH interferometer. The 2DEG in the
QH effect regime is confined to a Corbino disk region, shown by grey shadow. 
The inner and outer edge of the 2DEG are circles of radii $r_D$
and $r_U$, respectively. At the points $\xi_U$ and $\xi_D$ the Corbino disk is connected
to Ohmic reservoirs. The location of the inner Ohmic contact is such that the particular
interferometer shown here here is of the Mach-Zehnder type. If, in contrast, the inner
Ohmic contact is attached to the upper part of the Corbino disk, then such an interferometer
belongs to a Fabry-P\'{e}rot type.  
The quasi-particle tunneling between the inner and the outer edge takes place
at two QPCs, shown by dashed lines, that connect the points $\xi_{\ell}$ and $\xi'_\ell$, with
$\ell = L,R$.}
\vspace{-3mm}
\label{mz-scr}
\end{center}\end{figure}\end{center}

In view of various shortcomings of previous theoretical arguments, 
we propose to reconsider the physics of QH interferometers at the microscopic level. 
More specifically, we 
describe a microscopic model of a QH interferometer with the topology of a Corbino disk, 
schematically shown in Fig.\ \ref{mz-scr}.
The state of a QH liquid at filling factor $\nu = 1/m$, confined to a region between two circles of
radii $r_U$ and $r_D$, is described by Laughlin-type wave functions 
(\ref{laugh}) and (\ref{w-ring}).
The inner and outer edges of the QH liquid are connected to Ohmic
contacts at two points, $\xi_U$  and $\xi_D$, via strong electronic tunneling. 
Weak quasi-particle backscattering at two quantum point contacts (QPCs) is 
indicated in Fig.\ \ref{mz-scr} by dashed lines. It is described by the overlap of 
Laughlin states modified through the addition of quasi-holes located at
points $\xi_L$ and $\xi'_L$, for the left QPC, and at points $\xi_R$ and $\xi'_R$, for the right QPC. 
The position of the Ohmic contacts relative to QPCs determines whether the interferometer 
is of  MZ or FP type, as illustrated in Fig.\ \ref{mzfp} below.
We generalize the Laughlin wave function (\ref{laugh}) and (\ref{w-ring}) in order to take 
into account the deformation of 
the QH edges caused by the modulation gate and the Ohmic contacts. The same modification can 
be applied to describe more realistic QPCs, for which $\xi_L\simeq\xi'_L$
and $\xi_R\simeq\xi'_R$. 

We then use deformed Lauglin states to construct a 
subspace of low-energy excitations of a QH liquid.
The microscopic wave function (\ref{wave-1}) and (\ref{omega}),
resulting from weak incompressible deformations, is parameterized by 
an infinite set of variables
$t_k$. We invoke the classical plasma analogy  \cite{Laugh} and follow the steps of Ref.\  [\onlinecite{micro}]
in order to restrict the microscopic Hamiltonian $H$ to the subspace of these deformations. 
The restricted effective Hamiltonian is defined as ${\cal H} = PHP$, with $P$ being projector onto the low-energy subspace. After
the restriction, the variables $t_k$ turn into oscillator operators $a_k$ with canonical
commutation relations (\ref{commut-mic}).
These operators describe {\em gapless} plasmon excitations at the edges of the QH liquid.
The restricted Hamiltonian (\ref{h-edge}) contains an oscillator part 
with a linear spectrum and a Coulomb charging
energy, which depends on the number of quasi-particles, $M$, and the number of electrons, $N$, in the
QH system. The restricted tunneling operators take the form of vertex operators (\ref{h-tunnel}).
They are well defined single-valued operators, in contrast to quasi-particle operators.  
We do not find any trace of additional Klein factors \cite{Safi} in the tunneling 
amplitudes.

The low-energy theory so derived agrees with the effective theory of QH edge states, 
\cite{Wen,Frol} 
generalized to take into account the finite size of a QH system, the non-trivial topology
of a Corbino disk and the effects of a modulation gate and of a singular magnetic flux. 
The effective theory arises as a boundary contribution to the topological Chern-Simons field theory 
\cite{CS-theory} 
of a QH liquid in the bulk of a 2DEG in the presence of an external electromagnetic 
field.
In the context of this theory, our microscopic results for the relative phase 
of the tunneling amplitudes taken at different spatial points acquire a simple interpretation:
This phase corresponds to the insertion of a Wilson loop along the interferometer contour (\ref{eff-phas}). 
There are two 
contributions to this phase: one is the contour integral of the gauge
fields, the other one comes from the charge accumulated at the edges.

Having constructed the low-energy theory of an isolated QH system, we proceed to the analysis 
of the quasi-particle transport through the interferometer in response to the voltage bias applied 
to the Ohmic contacts. 
In our model, the Ohmic reservoirs have large capacitances. They therefore perfectly screen 
edge charges. In other words, any variation of zero modes, gate voltages, or the magnetic flux 
through the interferometer leads to an accumulation of charge at the Ohmic contacts, which is described by a local deformation of the edge. 
This property of Ohmic contacts agrees with the observation \cite{Averin-ohmic} that at the level of the effective theory, 
strong coupling to Ohmic contacts leads to a local elongation of the edge states near the Ohmic contacts. 
We treat electron- and quasi-particle tunneling perturbatively in order to describe rare transitions that change 
the numbers $N_D$ and $N_U$ of electrons at the inner and outer edge of the Corbino disk, respectively,
and the number $l = 0,\ldots, m-1$ of quasi-particles at the inner edge. 
The dynamics of the interferometer on a long time scale 
is described by a master equation for the probability of finding the system in 
a state with given values of the numbers $N_D$, $N_U$, and $l$.
Strong coupling of edge modes to Ohmic contacts, as opposed to quasi-particle tunneling, leads to the 
equilibration of inner and outer edge states at prescribed electro-chemical potentials.

By solving the master equation in the quasiparticle sector we finally find the
stationary current through the interferometer as a function of the singular magnetic flux $\Phi$
and the flux $\Phi_G$ induced by the modulation gate voltage. 
We thereby establish the {\it main result} of our paper: We find that quasi-particle tunneling 
rates in a QH interferometer with MZ topology (see Fig.\ \ref{mzfp}) 
depend on the magnetic flux 
through the combination $(\Phi+\Phi_G)/\Phi_0+l$. 
Thus, after summation over $l$, the AB oscillations in the stationary current 
acquire the electronic periodicity simply because a shift of the flux $\Phi$ by one flux quantum, $\Phi_0$,
is compensated by a shift of $l$ by 1. 
In contrast, for a QH interferometer with the  FP topology, tunneling rates are independent 
of the singular flux $\Phi$ and the number $l$. Therefore the stationary current oscillates as 
a function of $\Phi_G$ with the 
quasi-particle period $m\Phi_0$. In both cases the Byers-Yang argument holds true, and the 
paradox is resolved.

Our paper is organized as follows. We start by recalling the effective theory of the QH
effect in Sec.\ \ref{bulk-edge}. In Sec.\ \ref{s-wave-func}, we construct the ground
state wave function of an electronic interferometer in the fractional QH effect regime. 
In Sec.\ \ref{s-project}, we derive the low-energy theory 
of the interferometer by restricting the microscopic Hamiltonian and quasi-hole wave 
functions to the subspace of states corresponding to small incompressible 
deformations of a QH liquid. 
In Sec.\ \ref{conseq}, we propose a model of Ohmic contacts, study the effects 
of a singular magnetic flux and of a modulation gate, and develop the kinetic theory 
of an open QH interferometer out of equilibrium.

\section{Effective theory of the QH effect}
\label{bulk-edge}

For the purpose of comparison with the microscopic theory, 
we briefly recall, in this section, elements of a widely used effective theory of the QH effect. 
This theory provides a description of the low-energy physics of a 
QH state in the bulk of a 2DEG as well as of the edge states. In view of the universality of the low-energy
physics, and quite similarly to, e.g., the Landau-Ginzburg theory of superconductivity, the effective theory
of the QH effect is derived from general properties of a QH liquid that greatly reduce the number of relevant models.\cite{Wen,Frol} 
Those properties are the following ones:
(i) {\em Gauge invariance.} The effective model of edge states has to be chosen in such 
a way as to cancel the gauge anomaly in the action for the bulk of the 2D electron system (this corresponds to the conservation of
the total electric current, i.e., of the sum of the bulk current and the edge current, which are not conserved separately). 
(ii) {\em Existence of an electron operator.} Since, at the microscopic
level, the quantum Hall state is described by an electron wave function, 
there must exist at least one local operator of the effective theory describing the creation or annihilation of an electron; 
this operator transfers a charge $e$ and obeys Fermi statistics. 
(iii) {\em Single-valuedness of the wave function.} Because a QH state describes electrons, 
its wave function must be single-valued in the electron positions, irrespectively 
of whether quasi-particles are present. 
As a consequence, the mutual statistical phase of a quasi-particle and an electron must be an integer multiple of
$\pi$.  \cite{footnote2-mzf} Below we focus on a particular case of a QH system at filling factor $\nu=1/m$
and demonstrate how these properties serve to construct an effective theory for the bulk of 2DEG, 
and, consequently, an effective theory of edge states. The Appendix \ref{app-quant} contains a reminder 
on the quatization of the Chern-Simons theory.

\subsection{Action for the bulk of a QH liquid}

We start by considering an infinitely extended incompressible QH liquid, and address boundary
effects later, in Sec.\ \ref{boundary}. We assume that, in a QH liquid, there is 
a conserved current, $J^\mu$, where the index $\mu=0,1,2$ enumerates time and spatial components. 
The continuity equation, $\partial_\mu J^\mu = 0$, may be solved by introducing
a vector potential, $B_\mu$, with:
\begin{equation}
J^{\mu} = \frac{1}{2\pi}\epsilon^{\mu\nu\lambda}\partial_\nu B_\lambda.
\label{j-bulk}
\end{equation}
We will always use units where $e=\hbar=1$ and adopt the Einstein summation convention,
unless specified otherwise. The current is invariant under the gauge transformations
\begin{equation}
B_\mu \to B_\mu + \partial_\mu \beta.
\label{transf-app}
\end{equation}
This means that the action, $S[B]$, of the field $B_\mu$ 
must also be gauge-invariant. It must therefore have the form
\begin{multline}
S[B] = \alpha_0\int d^3r \epsilon^{\mu\nu\lambda}B_\mu\partial_\nu B_\lambda \\ + \alpha_1\int d^3r \partial_{[\mu} B_{\nu]}\partial^{[\mu} B^{\nu]} + \ldots .
\label{full-act}
\end{multline}
It is easy to see that the first term in the above action, the so called Chern-Simons
action 
\begin{equation}
S_0[B] = \alpha_0\int d^3r \epsilon^{\mu\nu\lambda}B_\mu\partial_\nu B_\lambda,
\label{bulk-act}
\end{equation}
has scaling dimension zero. By contrast, the (Maxwell-like) second term, as well as all other possible local terms 
not displayed in Eq.\ (\ref{full-act}), have negative scaling dimensions, i.e., they
are irrelevant at large distance and low energy scales.
Note that the action (\ref{bulk-act}) breaks time reversal symmetry, which is not prohibited 
for a system exposed to an external magnetic field.
It is also important to mention that the action (\ref{bulk-act}) is {\it topological}, 
i.e., it does not depend on the metric of the manifold on which it is defined 
(see Appendix \ref{app-quant} for details).

Next, the coupling of the current $J^\mu$ 
to an external electromagnetic field, described by a vector potential $A_\mu$, is given by the term:
\begin{equation}
S_{\rm int}[A,B] = \int d^3r A_\mu J^\mu
= \frac{1}{2\pi}\int d^3r A_\mu \epsilon^{\mu\nu\lambda}\partial_\nu B_{\lambda}.
\end{equation}
Integrating out the field $B_{\mu}$, we find the effective action for
the electromagnetic field. Neglecting irrelevant terms in $S[B]$, it has the Chern-Simons form:
\begin{equation}
S_{\rm eff}[A] = \frac{1}{(4\pi)^2\alpha_0} \int d^3r  \epsilon^{\mu\nu\lambda}A_\mu\partial_\nu A_\lambda.
\end{equation}
The average current, $\langle J^\mu\rangle = \delta S_{\rm eff}[A]/\delta A_\mu$, is then given by
Hall's law:
\begin{equation}
\langle J^\mu\rangle =
\sigma_H
\epsilon^{\mu\nu\lambda}\partial_\nu A_\lambda,
\end{equation}
where $\sigma_H=1/(8\pi^2\alpha_0) $ is the Hall conductivity.
Thus, we conclude that for an infinitely extended QH liquid the action (\ref{bulk-act}) correctly 
describes the QH current.

Having constructed the gauge-invariant low-energy action for an incompressible QH liquid, 
we proceed to an analysis of the spectrum of local excitations. 
It has been demonstrated \cite{CS-theory} 
that all states 
of the topological Chern-Simons theory with action (\ref{bulk-act}) 
are described by Wilson lines (see Appendix \ref{app-quant}). 
For instance, an operator annihilating a local excitation at the point $r$ 
and creating it at point $r'$ has the following form:
\begin{equation}
{\cal W}_q(r,r') = \exp\Big(i q\!\int^r_{r'}dr^\mu B_{\mu}\Big),
\label{op-bulk}
\end{equation}
where $q$ is a constant.
In order to demonstrate this, we note that the charge operator $Q_{\rm em}$ is given by the integral 
of the charge density $ J^0=(1/2\pi)\epsilon^{\nu\lambda}\partial_\nu B_\lambda$ 
over a space-like region enclosed by some contour $\gamma$. By Stokes' theorem it may therefore be written as
the contour integral
\begin{equation}
Q_{\rm em} =  (1/2\pi)\!\int_\gamma dr^\mu B_{\mu}.  
\label{charge}
\end{equation}
One can show that the commutator of these two operators is given by $[Q_{\rm em},{\cal W}_q]=\pm q{\cal W}_q/(4\pi\alpha_0)$, 
if $\gamma$ encloses only one of the two points $r$ and $r'$, and it vanishes otherwise. Therefore, the operator
${\cal W}_q$ creates two local excitations of charges $q/(4\pi\alpha_0)$ and $-q/(4\pi\alpha_0)$.
For the details of the derivation, see Appendix \ref{app-quant}.

Next, we note that the statistical phase $\theta_{12}$ 
of two local excitations 
(\ref{op-bulk}) with charges $q_1/(4\pi\alpha_0)$ and $q_2/(4\pi\alpha_0)$,
is determined by braiding the corresponding Wilson lines, \cite{CS-theory} 
and is equal to $\theta_{12} = \pi q_{1}q_{2}/4\pi\alpha_0$ (see Appendix \ref{app-quant} for the derivation).  
We now invoke constraint (ii), i.e., that the effective theory 
must contain an operator creating (or annihilating) an electron. This implies that the operator (\ref{op-bulk}) 
with $q = 4\pi\alpha_0$ (i.e., with unit charge) must obey the Fermi statistics. 
Thus, we require that $\pi(4\pi\alpha_0)^2/4\pi\alpha_0 = \pi m$, where $m$ is an odd integer.
This implies that $\alpha_0 = m/4\pi$, and 
hence the Hall conductivity takes the value $\sigma_H=1/8\pi^2\alpha_0  = 1/2\pi m$, in units where
$e=\hbar=1$. This characterizes the so called Laughlin series of FQHE states corresponding to a filling factor $\nu=1/m$, 
with $m$ being odd integer.

Apart from an electron with $q =m$, there are other local operators (\ref{op-bulk}) with a different
value of $q$.
Corresponding states must have single-valued wave-functions. In the context of effective theory, this implies
a statistical phase given by an integer multiple of $\pi$, when braided with an electron. When $\nu=1/m$
this leads to the condition that the number $q$ must be an integer.
Interestingly, the excitations created by the operator (\ref{op-bulk}) with $q = 1$ have fractional charge $Q_{\rm em} = 1/m$ 
and so-called fractional statistics, i.e., their statistical phase is $\theta = \pi/m$. These 
local excitations describe Laughlin quasi-particles. Thus, we conclude that, starting from the assumption
that the QH liquid supports only one conserved current we arrive at the effective theory of the 
QH effect at filling factors $\nu = 1/m$ only. Assuming that there exist several separately conserved currents, one arrives
at effective theories of QH states at more general filling factors.\cite{our-frac}

\subsection{Action of the edge degrees of freedom}
\label{boundary}

When a QH liquid at filling factor $\nu=1/m$ is confined to a finite region $D$,
the total effective action in the presence of an external electromagnetic field,
\begin{multline}
S[A,B]=S_0[B]+S_{\rm int} [A,B]\\ = \frac{1}{4\pi }\int_{D\times\mathbb{R}}d^3r\epsilon^{\mu\nu\lambda}\Big[2A_\mu+ mB_\mu\Big]\partial_\nu B_\lambda,
\label{s-bulk}
\end{multline}
is not-gauge invariant. One easily sees that, under a gauge transformation
\begin{equation}
A_\mu\to A_\mu + \partial_\mu\alpha,\quad B_\mu\to B_\mu + \partial_\mu\beta,
\label{gtr}
\end{equation}
 the action (\ref{s-bulk})
transforms as $S[A,B]\to S[A,B]+\delta S[A,B]$ with
\begin{multline}
\delta S[A,B] = \frac{1}{4\pi}\int_{D\times\mathbb{R}}d^3r \epsilon^{\mu\nu\lambda}
\Big[2\partial_\mu \alpha+m\partial_\mu\beta \Big]\partial_\nu B_\lambda \\
= \frac{1}{4\pi}\int_{\partial D\times\mathbb{R}}d^2r [2\alpha+m\beta]\epsilon^{\mu\nu}\partial_\mu b_\nu,
\label{anomal-ig}
\end{multline}
where $b_\nu$ is the restriction of the bulk field $B_\nu$
to the boundary $\partial D$. The physical reason for this gauge anomaly is found in the fact that, 
in a QH liquid confined to a finite region, the bulk Hall current (\ref{j-bulk}) is not conserved. 
Consequently, the electric charge may be accumulated at the edge of the system.

In order to restore the gauge invariance of the effective theory, we must take into account
boundary degrees of freedom. It is easy to see that the simplest boundary action
\begin{equation}
S[\phi] = \frac{m}{4\pi}\int_{\partial D \times\mathbb{R}} d^2r
[D_t\phi D_x\phi - h(D_x\phi) + \epsilon^{\mu\nu}b_\mu\partial_\nu\phi],
\label{s-edge1}
\end{equation}
where the coordinate $x$ parametrizes the boundary, cancels the gauge anomaly (\ref{anomal-ig}) 
if one assumes that the edge field $\phi$ transforms as
\begin{equation}
\phi\to\phi - (2/m)\alpha  - \beta,
\label{gaugetransform}
\end{equation}
and the covariant derivative is given by
the expression
\begin{equation}
D_\mu \phi = \partial_\mu\phi + (2/m)a_\mu + b_\mu,
\end{equation}
where $a_\mu$ is the restriction of $A_\mu$ to the boundary.

Note that the commutation relations for the field $\phi(x)$ are determined by the
first term in the action (\ref{s-edge1}):
\begin{equation}
[\partial_x\phi(x),\phi(y)] = \frac{2\pi i}{m}\delta(x-y).
\label{commut-eff}
\end{equation}
However, the precise form of the boundary Hamiltonian density, $(m/4\pi)h(D_x\phi)$, is not
fixed by the effective theory. The only requirement is that the Hamiltonian density 
should be a positive-definite function of $D_x\phi$. The simplest possible expression, 
$h = v(D_x\phi)^2$, justified if there are smooth confining potentials at the boundaries,
yields the Hamiltonian 
\begin{equation}
{\cal H} = \frac{mv}{4\pi}\int dx(D_x\phi)^2,
\label{ham}
\end{equation}
and leads to  
chiral edge modes with a linear dispersion law, as follows from the equation of motion.
Finally, the expression for the charge density at the edge of a QH system may be found by evaluating
the derivative $\rho= -\delta S/\delta a_t$ of the total action with respect to the boundary field. The result is:
\begin{equation}
\rho = \frac{1}{2\pi}D_x\phi.
\label{rho-edge}
\end{equation}

Next, we analyze the spectrum of local excitations at the edge. It is natural to assume
that such excitations are created by the local operator (\ref{op-bulk}), with a 
Wilson line starting and terminating at the boundary $\partial D$ of the QH system. 
Note, however, that this operator is not gauge-invariant. Similarly to 
the gauge anomaly in the action $S[A,B]$, this problem can be fixed by taking
into account the edge degrees of freedom. In particular, the gauge-invariant operator
that creates a Laughlin quasi-particle and a quasi-hole at the edge of a QH system
is given by
\begin{multline}
e^{i\phi(x)}{\cal W}_1(\xi,\xi')\exp\Big(i/m\!\int^{\xi}_{\xi'} dr^\mu A_{\mu}\Big)e^{-i\phi(x')} \\= 
e^{i\phi(x)}\exp\Big(i\!\int^{\xi}_{\xi'}
dr^\mu \big[B_{\mu}+\frac{2}{m}A_\mu\big]\Big)e^{-i\phi(x')},
\label{tun-ef}
\end{multline}
where the 1D coordinates $x$ and $x'$, and 2D coordinates $\xi$ and $\xi'$ are the different
parameterizations of the boundary $\partial D$, $\xi \propto e^{ix}$. The gauge invariance of the operator (\ref{tun-ef})
may be checked by carrying out the transformations (\ref{gtr}) and (\ref{gaugetransform}) on the r.h.s.\ of (\ref{tun-ef}).

We are particularly interested in an operator (\ref{tun-ef})
describing tunneling of a Lauglin quasi-particle from one edge to another one. In this case,
the tunneling points $\xi$ and $\xi'$, are very close to each other: $\xi\to \xi'$. Then the integrals
in the operator (\ref{tun-ef}) vanish, and denoting the resulting tunneling operator with ${\cal A}_{\rm qp}(\xi)$,
we write:
 \begin{equation}
 {\cal A}_{\rm qp}(\xi)=
e^{i\phi(x)}e^{-i\phi(x')}.
 \label{tun-ef-A}
 \end{equation}
The operator (\ref{tun-ef-A})
creates a pair of local charges at the boundary $\partial D$ of values $1/m$ and $-1/m$. 
This can be checked by evaluating the commutator of this operator with the charge density
operator (\ref{rho-edge}) with the help of Eq.\ (\ref{commut-eff}). It then becomes obvious,
that the operator annihilating an electron at the edge can be written as 
\begin{equation}
 \psi_{\rm el}(x) = \exp[im\phi(x)].
\label{el-eff}
\end{equation}
This operator has charge $1$ and Fermi statistics.

Finally, it is instructive to expand the edge field $\phi$ in oscillator modes.
Considering, for a moment, a single edge, we denote its length by $W$, set $a_\mu=b_\mu=0$,
and apply periodic boundary conditions on the field. We then arrive at the expression:
\begin{multline}
 \phi(x) = -\phi_N/m + 2\pi N x/W \\ + \sum_{k>0} \sqrt{\frac{2\pi}{kW}}\,[a_k e^{ikx} + a^\dag_k e^{-ikx}],
\label{ex-modes}
\end{multline}
where the creation and annihilation operators for plasmon modes satisfy the commutation relations 
$[a_k, a_{k'}^\dag] = (1/m)\delta_{kk'}$, and the zero modes satisfy $[N,e^{i\phi_N}] = e^{i\phi_N}$. 
Substituting this expansion
into the edge Hamiltonian (\ref{ham}), we find that
\begin{equation}
\mathcal{H} = \pi v m N^2/W  + m\sum_{k>0} vk a^\dag_ka_k .
\label{ham-eff-2}
\end{equation}
Thus, we conclude that the edge Hamiltonian indeed describes chiral edge plasmon modes with linear dispersion.
The effective model of edge states with Hamiltonian (\ref{ham-eff-2}), i.e., a chiral conformal field theory,
successfully describes several recent experiments; (see, e.g., Ref.\ [\onlinecite{our}]). 
However, when naively applied to MZ interferometers in the fractional
QH regime, it appears to lead to the Byers-Yang paradox, as mentioned in the introduction.

Indeed, let us consider the gedanken experiment illustrated in the upper panel of Fig.\ \ref{frol-simpl}.
In this formulation, the interferometer consists of two QH edges of infinite length coupled at two QPCs 
located at the points $\xi_L$ and $\xi_R$. 
We denote the boson field describing the upper and lower arms of the interferometer by $\phi_U(x)$ and $\phi_D(x)$, respectively.
The system shown in Fig.\ \ref{frol-simpl} is open. Therefore one can neglect zero modes in
the expansion (\ref{ex-modes}).
Next, we describe quasi-particle tunneling using the tunneling operators 
\begin{equation}
{\cal A}_{\rm qp}(\xi_\ell) = e^{i\phi_D(x_\ell)}e^{-i\phi_U(x'_\ell)},
\label{tt-e}
\end{equation}
where $\ell = L, R$. To leading order,
the oscillating contribution to the current through the interferometer 
is then given by $I = 2{\rm Re}\int dt \langle[{\cal A}_{\rm qp}^\dag(\xi_L,t), {\cal A}_{\rm qp}(\xi_R,0)]\rangle$.
Using Eq.\ (\ref{rho-edge}) and integrating out the field $B_\mu$, we find that 
the phase of the oscillating term is given by $(1/m)\int_\gamma A_\mu dx^\mu$, where $\gamma$ is the contour around the interferometer. 
In a situation, where the singular flux $\Phi$ is threaded through the interferometer, we find that
\begin{equation}
{\cal A}_{\rm qp}(\xi_L)^\dag {\cal A}_{\rm qp}(\xi_R) \propto 
\exp \left( \frac{2\pi i\Phi}{m\Phi_0}\right),
\end{equation}
i.e., the period of AB oscillations is $m\Phi_0$, in contradiction with the Byers-Yang theorem. 
Apparently, the problem with the naive approach arises because it considers an MZ interferometer
as an infinite system. In order to resolve the paradox, we need to consider a realistic,
finite QH systems, even if it is strongly coupled to ohmic reservoirs, as shown in the lower
panel of Fig.\
\ref{frol-simpl}. In what follows, we address the 
problem at the the microscopic level and then compare the results [see Eq.\ (\ref{compare})] to the
predictions of the effective theory.

\section{Microscopic description of a QH interferometer}
\label{s-wave-func}

In this section we construct the many-particle wave functions of the
ground state and of gapless excited states of an {\em isolated} QH interferometer with the Corbino 
disk topology. We then extend our results to the case of an {\em open} interferometer, connected to Ohmic reservoirs; 
see in Sec.\ \ref{conseq}. 
Our analysis proceeds step by step, starting from the
Laughlin wave function and manipulating it, with the purpose of identifying a realistic model of
interferometers. We first present the most important steps and then make them precise in Sec.\ \ref{s-incompr},
using the classical plasma analogy.  \cite{Laugh} The variational wave function, proposed by Laughlin in
Ref.\  [\onlinecite{Laugh}] and later justified by Haldane and Rezayi in Ref.\  [\onlinecite{Haldane}], 
describes an approximate ground state,
$|N\rangle$, of a QH system with $N$ electrons at filling factor $\nu = 1/m$:
\begin{equation}
\langle \underbar{z}\,|N\rangle = \prod_{i<j}^N(z_i-z_j)^m \exp\Big( -\sum_i^N \frac{|z_i|^2}{4l_B^2}\Big).
\label{laugh}
\end{equation}
Here $\underbar{z}$ denotes a set of complex coordinates $z_i = x_i + iy_i$
describing the position of the $i^{\rm th}$ electron, $i = 1\dots N$, and $l_B = \sqrt{\hbar c/eB}$
is the magnetic length. It is known  \cite{QHE} that the wave function (\ref{laugh})
describes a circular droplet of a QH liquid of constant density $\rho_{\rm bg} = 1/(2\pi ml_B^2)$
and of radius $r = l_B\sqrt{2mN}$.

In the next step, we add to the state (\ref{laugh}) a macroscopic number, $M$, of Laughlin quasi-particles
 \cite{Laugh} at the origin:
\begin{equation}
\langle \underbar{z}\,|N,M\rangle = \prod_i^N z_i^M \langle \underbar{z}\,|N\rangle.
\label{w-ring}
\end{equation}
In Sec.\ \ref{s-incompr} we explicitly show that the wave function (\ref{w-ring})
describes a QH state of constant electron density $\rho_{\rm bg}$ inside a Corbino disk,
as shown in Fig.\ \ref{mz-scr}. The inner hole of the disk has
a radius  $r_D = l_B\sqrt{2M}$, while the outer radius of the disk is given by
$r_U = l_B\sqrt{2(M+mN)}$.

Additional small incompressible deformations of the QH liquid may be described as follows.
We note that, in a suitable gauge, all states of the lowest Landau level can be described by holomorphic
functions of the electron coordinates. We therefore look for a wave function of the form  \cite{micro, fr-icm}
\begin{equation}
\langle \underbar{z}\,|N,M,\underbar{t}\,\rangle = \exp\Big[ m\sum_i^N 
\omega(z_i)\Big]\langle \underbar{z}\,|N,M\rangle,
\label{wave-1}
\end{equation}
where the function
\begin{equation}
\omega(z) = \sum_{k}t_kz^k
\label{omega}
\end{equation}
is analytic inside the Corbino disk (shown in Fig.\ \ref{mz-scr}), and $\underbar{t}$ denotes a set of
parameters $t_k$, $k\in \mathbb{Z}$. In Sec.\ \ref{s-incompr} we show that the shape of the deformed disk is
given by the solution of a two-dimensional electrostatic
problem, with $\omega(z)$  playing the role of an external potential.

We utilize the wave function (\ref{wave-1}) in two ways. First of all,
small incompressible deformations are known to be the gapless excitations of the QH
state.  \cite{QHE} Thus, in order to describe the low-energy physics of a QH liquid,
we determine matrix elements of the microscopic Hamiltonian between wave functions 
(\ref{wave-1}) describing incompressible deformations.
Second, we investigate the effects of a modulation gate located near one of the arms
of the interferometer and of two Ohmic contacts. Both effects  may be
described by local edge deformations: the charge expelled by the
modulation gate or added to the interferometer, is eventually  
concentrated at the Ohmic contacts. Because of the linearity 
of the resulting equations in the case of weak deformations,  the spectrum 
of excitations and the properties of the ground state of a realistic model of a
QH interferometer may, at first, be studied 
separately. We use our results in Sec.\ \ref{conseq}, where we investigate 
the AB effect in the current
through a QH interferometer connected to Ohmic reservoirs.

\subsection{Plasma analogy and incompressible states}
\label{s-incompr}

The classical plasma analogy \cite{Laugh} has proven to be an efficient method in the analysis
of QH states.  \cite{Capelli,plasma-analog}
It relies on the important observation that the norm of the Laughlin wave function may be written as the
partition function of an ensemble of $N$ classical particles interacting via the two-dimensional (logarithmic)
Coulomb potential. In the large-$N$ limit, the evaluation of this partition function reduces to solving a
two-dimensional electrostatic problem. Here we apply this method directly to the wave function
(\ref{wave-1}). We write
\begin{multline}
Z=\int d^2z_1\ldots d^2z_N |\langle \underbar{z}\,|N,M,\underbar{t}\,\rangle|^2 
\\ = \int d^2z_1\ldots d^2z_N e^{- m E_{\rm pl}},
\label{Z1}
\end{multline}
where the inverse temperature of the plasma is $m$ and the energy is given by the expression:
\begin{multline}
E_{\rm pl} = -\sum_{i<j}\ln |z_i-z_j|^2  \\
+  \sum_i\bigg[\frac{|z_i|^2}{2ml_B^2}- \sum_i \frac{M}{m}\ln |z_i|^2 -
2{\rm Re}\,\omega(z_i)\bigg].
\label{E}
\end{multline}

Introducing the microscopic charge density $\rho(z)=\sum_i\delta^2(z-z_i)$, we can formally write
\begin{multline}
E_{\rm pl} = -\frac{1}{2}\iint\! d^2zd^2w\, \rho(z)\rho(w)\ln|z-w|^2 \\ - \int d^2z \rho(z)\varphi_{\rm ext}(z),
\label{ener}
\end{multline}
where
\begin{equation}
\varphi_{\rm ext}(z) = -\frac{|z|^2}{2ml_B^2}+\frac{M}{m}\ln |z|^2+2{\rm Re}\,\omega(z),
\label{extphi}
\end{equation}
and divergent self-energies are suppressed in the first term on the right hand side of Eq.\ (\ref{ener}). 
This representation makes it obvious that the partition function (\ref{Z1})
describes a gas of charged particles interacting via the 2D Coulomb potential and confined
by the external potential $\varphi_{\rm ext}(z)$.
The first term in Eq.\ (\ref{extphi})
describes the interaction of particles with a neutralizing homogeneous background charge
of density $\rho_{\rm bg}=(1/4\pi)\Delta (|z|^2/2ml_B^2) = 1/2\pi m l_B^2$. The second term
can be interpreted as describing repulsion caused by a macroscopic charge $M/m$ at the origin. Finally, the
last term describes the effect of an external (chargeless, since $\Delta {\rm Re}\,\omega(z) =0$)
potential on the particles in the gas.

The next step is to approximate the integral over coordinates in Eq.\ (\ref{Z1})
by a functional integral \cite{footnoterho} over the density $\rho(z)$:
\begin{equation}
Z= \int {\cal D}\rho(z) e^{- m E_{\rm pl}[\rho]}.
\label{Z2}
\end{equation}
In Eq.\ (\ref{Z2}), we neglect the Jacobian of the transformation from the variables $\underbar{z}$
to $\rho(z)$.\cite{Dyson} Taking the Jacobian into account is crucial for the correct 
description of the the physics at the microscopic length scale $l_B$, which, however,
is not the subject of the present work.  
In this approximation,   
the evaluation of $Z$ becomes straightforward.  We note that
the energy of the plasma is a quadratic function of the density. Hence the average density
$\langle\rho(z)\rangle=Z^{-1}\langle N,M,\underbar{t}\,|\rho(z)|N,M,\underbar{t}\,\rangle$
is given by the solution
of the saddle-point equation
$\delta E_{\rm pl}/\delta\rho(z) =0$, which reads
\begin{equation}
\int d^2w \langle\rho(w)\rangle\ln|z-w|^2 + \varphi_{\rm ext}(z) = 0
\label{eom-1}
\end{equation}
for the domain where $\langle\rho(z)\rangle\neq 0$.
Thus, in the large-$N$ limit, an ideal QH liquid may be described by solving a 2D electrostatics problem.\cite{crystal}

Among important consequences of this simple equation are the following ones. First, it implies that the total
potential vanishes in the region where $\langle\rho(z)\rangle\neq0$, i.e.\ where the 2DEG is not fully depleted.
In other words, the Coulomb plasma is a ``perfect metal'' that completely screens the
external potential $\varphi_{\rm ext}$. Applying the Laplacian to (\ref{eom-1}),
we find that $\langle\rho(z)\rangle=\rho_{\rm bg}$, i.e., the Coulomb plasma is distributed homogeneously
to screen the background charge. This confirms that the wave function (\ref{wave-1}) describes
an incompressible deformation of the QH droplet. In particular, the wave function (\ref{w-ring})
describes the approximate ground state of a QH interferometer with the shape of a Corbino disk. 
Indeed, the plasma
analogy suggests that the hole in the Corbino disk is formed symmetrically around the origin, 
where the macroscopic charge $M/m$ is located. It serves to screen this charge, so that the
total potential vanishes in the region occupied by the 2DEG. Because of perfect screening, the shape
of the outer edge is, however, independent of the shape and position
of the hole and displays the symmetry of boundary conditions in the background charge distribution 
(see the first term
in Eq.\ (\ref{extphi})).

\begin{center}\begin{figure}[htb]\begin{center}
\epsfxsize=5.5cm
\epsfbox{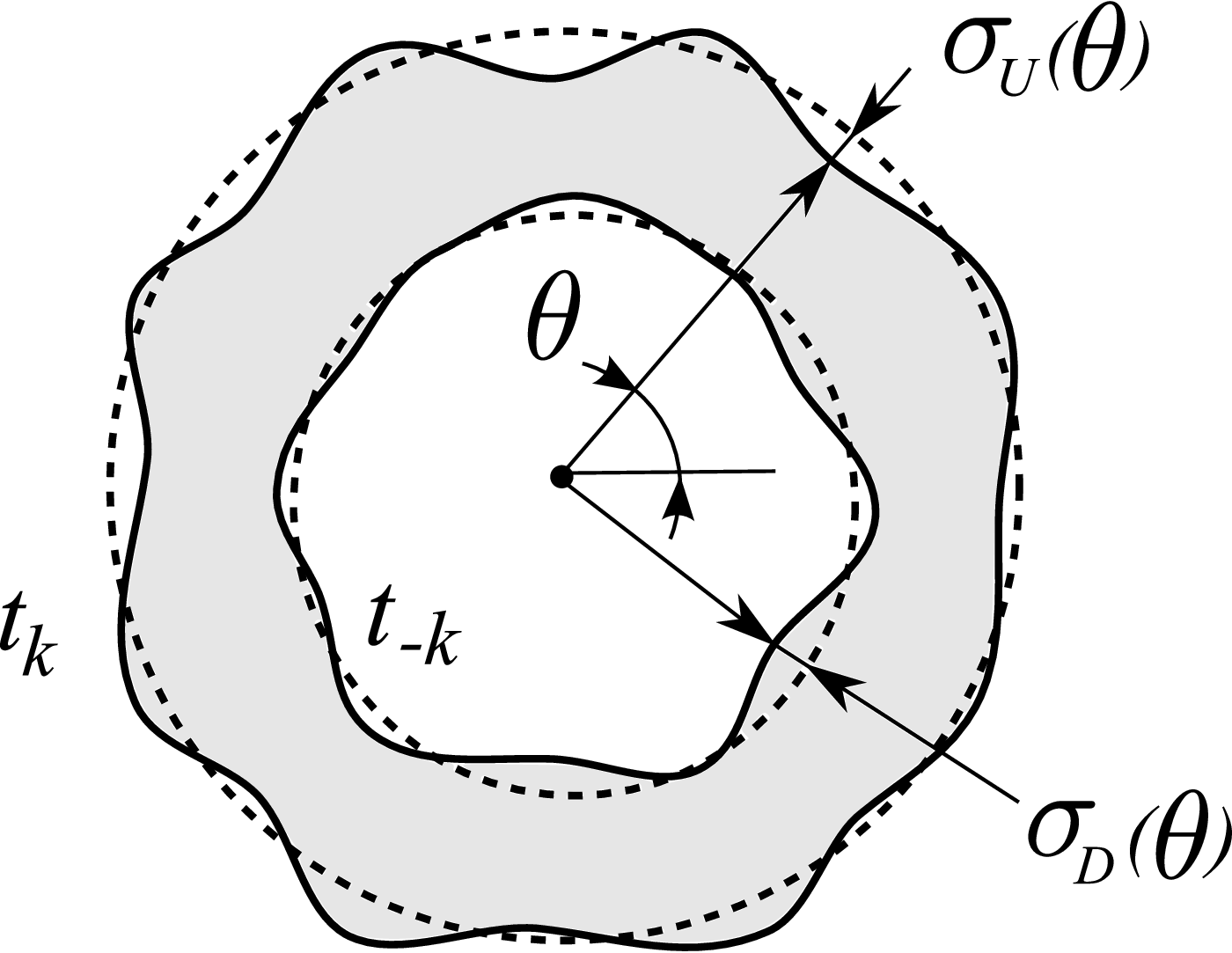}
\caption{A QH liquid, whose ground state is given by Eq.\ (\ref{wave-1}), is shown schematically.
The dashed lines show the edges of the unperturbed Corbino disk, while an incompressible
deformation is shown by the full lines. The region of constant electron density
$\langle\rho(z)\rangle=\rho_{\rm bg}$  is shown by the grey shadow. The shape of this region
is determined by the complex function $\omega(z)$ defined in Eq.\ (\ref{omega}). The Fourier
components of  charge density $\sigma_U$ accumulated at the outer edge are given by the
coefficients $t_k$ of the regular part of the Laurent series for $\omega(z)$, while the Fourier
components of  charge density $\sigma_D$ accumulated at the inner edge are given by the coefficients
$t_{-k}$ with $k>0$ of the singular part of $\omega(z)$ [see Eq.\ (\ref{sol-def})].}
\label{def}
\end{center}\end{figure}\end{center}

We now investigate the effect of the potential $\omega(z)$ perturbatively.
Let us denote by $D$ the region to which the QH system is confined. We
search for the solution of Eq.\ (\ref{eom-1}) in the form
$\langle\rho(z)\rangle = \rho_{\rm bg}$, for $z\in D$, and $\langle\rho(z)\rangle = 0$
otherwise. Thus we can rewrite Eq.\ (\ref{eom-1}) as
\begin{equation}
\rho_{\rm bg}\int_D d^2w \ln|z-w|^2 + \varphi_{\rm ext}(z) = 0.
\nonumber
\end{equation}
Considering a small deformation, $D = D_0 + \delta D$, and taking into account that
the integral over the undeformed Corbino disk, $D_0$, cancels the first two terms in
Eq.\ (\ref{extphi}), we arrive at the following result:
\begin{equation}
\rho_{\rm bg}\int_{\delta D} d^2w \ln|z-w| + {\rm Re}\,\omega(z) =0.
\label{eom-2}
\end{equation}
In polar coordinates (see Fig.\ \ref{def}), the boundaries of the deformed disk can
be parameterized by the function $r(\theta) = r_s\pm\sigma_s(\theta)/\rho_{\rm bg}$, where $s = U,D$, and
$\sigma_s(\theta)$ are 1D charge densities accumulated at the inner and outer edge
due to the deformation.

Because of perfect screening in the two-dimensional Coulomb plasma,
one can solve Eq.\ (\ref{eom-2}) independently for each edge.
Using the series expansion 
\begin{equation}
\ln(z-w) = \ln(z) - \sum_{k>0}(w/z)^k/k,\quad |z|\geq|w|,
\label{exp-n}
\end{equation}
and the explicit expression (\ref{omega}) for the potential $\omega(z)$,
we can solve Eq.\ (\ref{eom-2}) to first order in $\sigma_s$ by using power series. 
The result can be presented in the form
of Fourier series:
\begin{subequations}
\label{sol-def}
\begin{eqnarray}
2\pi r_U\sigma_U(\theta) &=& 2{\rm Re}\sum_{k>0}kt_kr_U^k e^{ik\theta}, \\
2\pi r_D\sigma_D(\theta) &=& 2{\rm Re}\sum_{k>0}kt_{-k}r_D^{-k} e^{-ik\theta}.
\end{eqnarray}
\end{subequations}
These series show how the microscopic wave function (\ref{wave-1})
determines the shape of the deformed Corbino disk.
\begin{center}\begin{figure}[tb]\begin{center}
\epsfxsize=7cm
\epsfbox{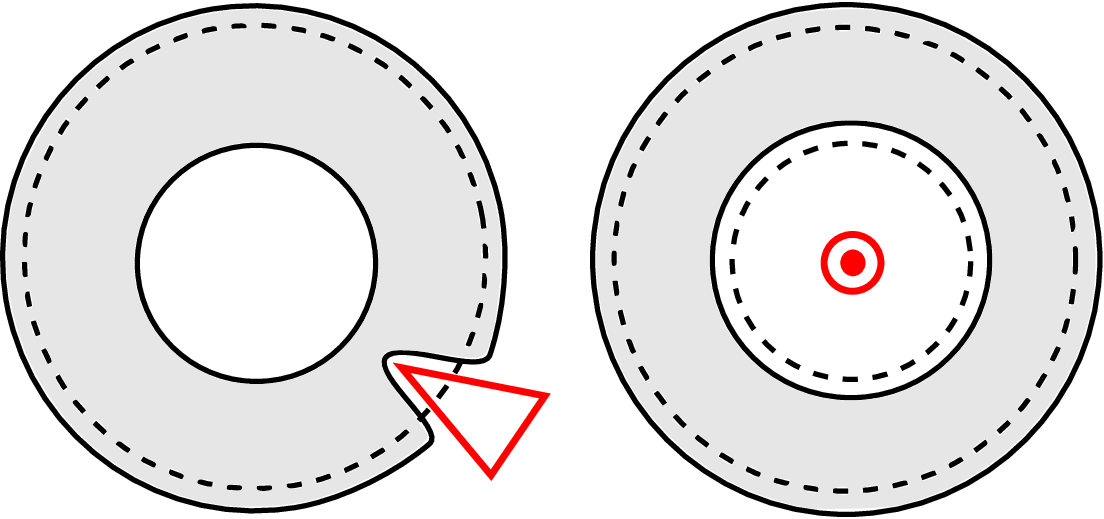}
\caption{The effects of a modulation gate and of a singular flux are illustrated. {\em Left panel}:
The modulation gate locally depletes the QH liquid. Due to the incompressibility of the QH liquid,
the repelled charge is accumulated homogeneously along the edge. {\em Right panel}:
In the language of the Coulomb plasma, the multiplier in the wave function
(\ref{w-magn}) depending on the singular flux can be viewed as a point-like charge placed in the
center of the inner hole. It homogeneously shifts both edges to preserve the electro-neutrality.}
\label{mod}
\end{center}\end{figure}\end{center}

Next, we analyze the effects of a modulation gate and of a singular magnetic flux on an {\it isolated} QH droplet,
as illustrated in Fig.\ \ref{mod}.
According to Eqs.\ (\ref{sol-def}), the deformation caused by the modulation gate can be described by the wave function
\begin{multline}
\langle \underbar{z}\,|N,M,\Phi_G\rangle   =\exp\Big[-\sum_i^N\frac{\Phi_G}{\Phi_0}f\Big(\frac{z_i}{\xi_G}\Big)\Big]
\langle \underbar{z}\,|N,M\rangle,
\label{w-modg}
\end{multline}
where we have introduced the function 
\begin{equation}
f(x) =  \sum_{k>0}x^k/k=-\ln(1-x).
\label{fx}
\end{equation} 
Indeed, the 1D charge accumulated at the edge as a result of the deformation has the following form:
\begin{equation}
r_U\sigma_U(\theta) =
-\frac{\Phi_G}{m\Phi_0}\Big(\delta(\theta-\theta_0)-\frac{1}{2\pi}\Big),
\label{dens-g}
\end{equation}
where $\theta_0$ is the argument of the position, $\xi_G$, of the modulation gate.
This function correctly captures the effects of a modulation gate on an isolated interferometer: 
the local depletion
of the 2DEG at the point $\xi_G = r_Ue^{i\theta_0}$ and the homogeneous
expansion of the QH liquid due to its incompressibility. This observation agrees with
the fact that the function (\ref{fx}) of the complex variable $x$ is electro-neutral from the 
2D electrostatics 
point of view, because it has two branch points, at $x=1$ and $x=\infty$.
Finally, it is easy to check that the flux through the depleted area under the modulation gate,
$-B\int (\sigma_U/\rho_{\rm bg})r_Ud\theta$, is indeed equal to $\Phi_G$.

After an adiabatic insertion of a singular magnetic flux $\Phi$
through the hole in the Corbino disk at the origin, the wave function
acquires the phase factor $\prod_i\exp[-i\Phi/\Phi_0\arg z_i]$.
Furthermore, the wave function is deformed by the ``spectral flow'' preserving its 
single-valuedness. This effect is described by
the additional multiplier $\prod_i z_i^{\Phi/\Phi_0}$. The overall effect of a singular flux on the
wave function (\ref{w-ring}) can thus be represented by replacing the original wave function by
\begin{equation}
\langle \underbar{z}\,|N,M,\Phi\rangle = \prod_i^N |z_i|^{\Phi/\Phi_0}\langle \underbar{z}\,|N,M\rangle.
\label{w-magn}
\end{equation}
Thus, in the presence of a singular magnetic flux, the plasma energy contains an additional term
\begin{equation}
\delta E_{\rm pl} = -\frac{\Phi}{m\Phi_0}\sum_i\ln|z_i|^2.
\label{E-sing}
\end{equation}
In the language of the Coulomb plasma it describes the addition of a charge $\Phi/m\Phi_0$ at the origin in
the hole of the interferometer. This shifts the edges of the
Corbino disk by an amount 
\begin{equation}
\delta r_s = \frac{\Phi}{m \Phi_0}\frac{1}{2\pi r_s\rho_{\rm bg}},\quad s = U,D,
\label{flux-shift}
\end{equation}
as illustrated in Fig.\ \ref{mod}.

\subsection{Low-energy subspace}
\label{s-low-energy}

Having found the set of states (\ref{wave-1}) describing incompressible deformations of a QH liquid,
we proceed to construct operators generating the subspace of such low-energy
states when applied to the undeformed ground-state and to identify their commutation relations.
First of all, we introduce zero-mode operators changing
the number of electrons, $N$, and of quasi-particles, $M$, in the system
\begin{subequations}
\begin{eqnarray}
e^{i\phi_M}|N,M,\underbar{t}\,\rangle &=& |N,M+1,\underbar{t}\,\rangle, \\
e^{i\phi_N}|N,M,\underbar{t}\,\rangle &=& |N+1,M,\underbar{t}\,\rangle.
\end{eqnarray}
\label{zeromodedef}
\end{subequations}
States corresponding different numbers of electrons $N$ are obviously
orthogonal. The orthogonality of states with different numbers of quasi-particles
$M$ follows from the fact that in a symmetric Corbino disk such states
have different angular momenta.   Taking this observation into account,
one derives from definitions (\ref{zeromodedef}) the following commutation relations
for the zero-mode operators: \cite{footnote-zeromodes}
\begin{equation}
[M,e^{i\phi_M}] = e^{i\phi_M}, \; \; [N,e^{i\phi_N}] = e^{i\phi_N}.
\label{zeromodecom}
\end{equation}
Next,  we introduce deformation operators
$a_{ks}$, $s = U,D$, $k>0$, acting on the right as
\begin{subequations}
\label{ak}
\begin{eqnarray}
a_{kU}|N,M,\underbar{t}\,\rangle &=& -i\sqrt{k}r_U^kt_k|N,M,\underbar{t}\,\rangle, \\
a_{kD}|N,M,\underbar{t}\,\rangle &=& -i\sqrt{k}r_D^{-k}t_{-k}|N,M,\underbar{t}\,\rangle.
\end{eqnarray}
\end{subequations}
Apparently, the states (\ref{wave-1}) are coherent states for these operators.
In order to find their commutation relations, we need to evaluate
scalar products of the states (\ref{wave-1}).

We start by calculating the norm of a wave function, which is given by the square root of the partition
function of the Coulomb plasma, (\ref{Z1}) and (\ref{Z2}), and evaluate the
``free energy''
\begin{equation}
F_{\rm pl}=-(1/m)\log(Z).
\label{freenergydef}
\end{equation}
Treating the potential $\omega(z)$
as a perturbation, we obtain
\begin{equation}
F_{\rm pl} = F_0 +\frac{\rho_{\rm bg}^2}{2}\iint_{\delta D} d^2zd^2w \ln |z-w|^2,
\nonumber
\end{equation}
where the constant $F_0$ is the contribution from the unperturbed state and from
the determinant of the Gaussian integral.
We evaluate the integral on the r.h.s.\ with the help of the solution (\ref{sol-def}) and present the result
as a bilinear form in the coefficients $t_k$,
\begin{multline}
F_{\rm pl}(\underbar{t}^*,\underbar{t}\,) = F_0  \\ -\sum_{k>0}k[ r_U^{2k}t^*_{k}t_{k}+r_D^{-2k}t^*_{-k}t_{-k}
+ t_kt_{-k}+t_k^*t_{-k}^*].
\label{free-en}
\end{multline}
The analytic structure of this bilinear form allows us to extend the result
for the norm $Z = \exp\{-mF_{\rm pl}(\underbar{t}^*,\underbar{t}\,)\}$ to the scalar products
\begin{equation}
\langle N,M,\underbar{t}\,|N,M,\underbar{t}'\rangle = \exp\{-mF_{\rm pl}(\underbar{t}^*,\underbar{t}')\}.
\label{products}
\end{equation}

Next, we define differential operators, via their matrix elements, as follows:
\begin{equation}
\langle\Psi|\frac{\partial}{\partial t_k}| N,M,\underbar{t}'\rangle
= \frac{\partial}{\partial t'_k}\langle \Psi|N,M,\underbar{t}'\rangle. 
\label{notstandart}
\end{equation}
A straightforward calculation then yields:
\begin{multline}
\langle N,M,\underbar{t}\,|\frac{\partial}{\partial t_k}|N,M,\underbar{t}\,'\rangle
\\ = mk[\, r_U^{2k}t^*_k  +t'_{-k}]
\langle N,M,\underbar{t}\,|N,M,\underbar{t}\,'\rangle.
\label{interm}
\end{multline}
Using definition (\ref{ak}), we may write 
\begin{multline}
\langle N,M,\underbar{t}\,
|a_{kU}^\dag|N,M,\underbar{t}\,'\rangle \equiv \langle N,M,\underbar{t}
\,'|a_{kU}|N,M,\underbar{t}\,\rangle^* \\ =i
\sqrt{k}r_U^kt^*_{k} \langle N,M,\underbar{t}\,|N,M,\underbar{t}\,'\rangle.
\label{matrixelements}
\end{multline}
Substituting this equation in Eq.\ (\ref{interm}), we find the expression for the adjoint operators 
acting on the states $|N,M,\underbar{t}\,\rangle$:
\begin{equation}
a_{kU}^\dag = \frac{ir_U^{-k}}{m\sqrt{k}}\Big(\frac{\partial}{\partial t_k}
- mkt_{-k}\Big).
\label{a-ku}
\end{equation}
Repeating the same calculations for the operators $\partial/\partial t_{-k}$, we obtain:
\begin{equation}
a_{kD}^\dag = \frac{ir_D^{k}}{m\sqrt{k}}\Big(\frac{\partial}{\partial t_{-k}}
- mkt_{k}\Big).
\label{a-kd}
\end{equation}

Using Eqs.\ (\ref{ak}), (\ref{a-ku}) and (\ref{a-kd}) for the operators $a_{ks}$ and their adjoints
and the relation $[\partial/\partial t_k, t_{k'}] = -\delta_{kk'}$,\cite{footnote-com}
 we obtain the commutation relations:
\begin{equation}
[a_{ks},a^\dag_{k's'}] = \frac{1}{m}\delta_{kk'}\delta_{ss'}.
\label{commut-mic}
\end{equation}
Similarly, one finds that $[a_{ks}^\dag,a_{k's'}^\dag] =[a_{ks},a_{k's'}] =0$. Thus
the operators $a_{ks}$ and $a_{ks}^\dag$ introduced in (\ref{ak}), (\ref{a-ku}) and (\ref{a-kd}) satisfy canonical 
commutation relations, and the subspace of incompressible deformations has a natural Fock space 
structure with respect to these operators.

\section{Restriction to the low-energy subspace}
\label{s-project}

Starting from the microscopic model, we now explicitly derive the low-energy effective theory
of an interferometer. For this purpose,  we restrict the microscopic Hamiltonian and
the overlaps of wave functions of quasi-particles at the two edges to the low-energy subspace
constructed above.  The restriction of these operators is defined by $O\to POP$,
where the orthogonal projection $P$ is given by:
\begin{multline}
P = \sum_{N,M}\int 
\prod_kd^2t_k\prod_kd^2t'_k  \\ \times
|N,M,\underbar{t}\,\rangle \langle N,M,\underbar{t}\,|N,M,\underbar{t}'\,\rangle^{-1} \langle N,M,\underbar{t}'\,|,
\label{projection-def}
\end{multline}
where
the inverse is defined in the sense of the inverse kernel.
We first implement the restriction procedure for a symmetric Corbino disk. 
Then, in Sec.\ \ref{conseq}, we take advantage of the fact that incompressible deformations are weak
and lead to linear relations (\ref{eom-2}) and (\ref{sol-def}). The oscillator
operators are shifted in order to take into account additional deformations caused by Ohmic contacts, modulation
gate, and a singular magnetic flux.

\subsection{Edge Hamiltonian}
\label{s-edge}

The microscopic Hamiltonian for $N$ electrons, restricted to the lowest Landau level, is given by the expression
\begin{equation}
H = \sum_{i}^NU(z_i)+ \sum_{i<j}^N V(|z_i-z_j|),
\label{microham}
\end{equation}
where $V(|z|)$ is the potential of the screened 3D Coulomb interaction and $U(z)$ is the
confining potential, which forces electrons to stay within the interferometer. Note that there is no 
kinetic energy operator in (\ref{microham}), because when acting on the lowest Landau level 
it gives a constant contribution,
$N\hbar\omega_c/2$, where $\omega_c = eB/m_ec$
is the cyclotron frequency. 

To find the restriction of the microscopic Hamiltonian to the subspace
of incompressible deformations, $\mathcal{H} = PHP$, one needs to evaluate the matrix elements
\begin{equation}
E(\underbar{t}^*,\underbar{t}') = \frac{\langle N,M,\underbar{t}\,|H
|N,M,\underbar{t}'\rangle}{\langle N,M,\underbar{t}\,|N,M,\underbar{t}'\rangle}.
\label{edge-energy}
\end{equation}
We first consider the diagonal matrix elements $E(\underbar{t}^*,\underbar{t})$ in (\ref{edge-energy}).
They can be rewritten in terms of the electron density in the deformed state as follows:
\begin{multline}
E(\underbar{t}^*,\underbar{t}\,) \simeq \int d^2z U(z)\langle\rho(z)\rangle
 \\ +\frac{1}{2}\iint d^2z d^2w V(|z-w|)\langle\rho(z)\rangle\langle\rho(w)\rangle,
\label{edge-energy-2}
\end{multline}
where we have applied the approximation
$\langle\rho(z)\rho(w)\rangle \simeq \langle\rho(z)\rangle\langle\rho(w)\rangle$,
neglecting the correlations or ``exchange contributions'', which is justified in the large-$N$ limit.

Next, we  express the restricted Hamiltonian 
in terms of the deformation operators (\ref{ak}). To this end, we consider small deformations
of the state (\ref{w-ring}) and take into account the fact that the density is constant,
$\langle\rho(z)\rangle = \rho_{\rm bg}$, for $z\in D$. Writing the deformed region as
$D = D_0 +\delta D$, one can expand the integral (\ref{edge-energy-2})
in the small deformation $\delta D$
and evaluate the correction term with the help of the result (\ref{sol-def}):
\begin{multline}
E(\underbar{t}^*,\underbar{t}\,) =E_0
+ m\sum_{k>0}\,[\varepsilon_U(k)(r_U)^{2k}kt^*_kt_k 
\\ +\varepsilon_D(k)(r_D)^{-2k}kt^*_{-k}t_{-k}],
\label{edge-spect}
\end{multline}
where $E_0$ is the energy of a QH system confined to an undeformed Corbino disk,
and the last two terms originate from the deformation $\delta D$.
The excitation spectra, $\varepsilon_s(k)$, $s = U,D$,
are determined by the 3D Coulomb interaction and by the confining potential:
\begin{multline}
\varepsilon_s(k) = \frac{kU'(r_s)}{2\pi m\rho_{\rm bg}r_s}
\\ +\frac{k}{m}\int_0^{2\pi}\!\!d\varphi V\Big(2r_s|\sin\frac{\varphi}{2}|\Big) [e^{ik\varphi}-e^{i\varphi}].
\label{spectr-scr}
\end{multline}
Using again the analytic structure of the bilinear form
(\ref{edge-spect}) to extend this result to off-diagonal matrix elements, we 
find the restricted Hamiltonian to have the following form:
\begin{equation}
\mathcal{H} = E_0+m\!\sum_{s=U,D}\sum_{k>0}\varepsilon_s(k)a_{ks}^\dag a_{ks}\,.
\label{h-edge1}
\end{equation}

We further assume that the potential $V$ describes Coulomb interactions
screened at a distance $d$.
In the low-energy limit, i.e., for $kd/r_s\ll 1$, the deformation energy
in (\ref{spectr-scr}) is then linear as a function of the mode number; i.e., $\varepsilon_s(k) \simeq v_s k/r_s$,
where the constants $v_s$ are the group velocities of edge excitations.
As a function of the number of electrons $N$ and the number of quasi-particles $M$,
the energy of the undeformed state takes the following form:
\begin{equation}
E_0(N,M) = \frac{v_D}{2mr_D}M^2+ \frac{v_U}{2mr_U}(M+mN)^2.
\label{ener-zero}
\end{equation}
Replacing the mode number by the wave vector, $k\to kr_s$,
we arrive at our final expression for the edge Hamiltonian:
\begin{equation}
\mathcal{H} = E_0(N,M)+m\!\sum_{s=U,D}\sum_{k>0}v_ska_{ks}^\dag a_{ks}\,.
\label{h-edge}
\end{equation}

We conclude this section with an important remark.
The right hand side of Eq.\ (\ref{spectr-scr}) contains two terms. The first one 
originates from the confining potential and the second one comes from Coulomb interactions.
Consequently, the velocities of edge excitations contain
two contributions:
\begin{equation}
v_s = cE(r_s)/B + (e^2/m\hbar)\ln(d/l_B).
\label{velo}
\end{equation}
The first term
is the drift velocity, which is proportional to the boundary electric field, $E(r_s)$,
while the second one is proportional to the ``Coulomb logarithm''.
The ultraviolet cutoff in (\ref{velo}) is determined by correction terms in a $1/N$-expansion
of the two-point density correlation function. In fact, Eq.\ (\ref{velo}) coincides with
an expression proposed earlier in Ref.\ [\onlinecite{our-frac}], in
the framework of the effective theory, on the basis of
the classical electrostatic picture.

\subsection{Tunneling Hamiltonian}
\label{s-tunnel}

The tunneling Hamiltonian of an interferometer may be written as a sum
of tunneling operators at the left and right QPC:
\begin{equation}
{\cal H}_T=\sum_{\ell=L,R}[ {\cal A}_{\rm qp}(\xi_{\ell})+ {\cal A}^\dag _{\rm qp}(\xi_{\ell})].
\nonumber
\end{equation}
The tunneling operator ${\cal A}_{\rm qp}(\xi)$ is an operator annihilating a quasi-particle at a point $\xi$ 
on one edge
and recreating it at a point $\xi'$ on the other edge. Note, that the tunneling path is typically short,
therefore we will take a limit $\xi'\to\xi$ in the end of calculations. 
At low energies, the tunneling operator can be defined as an operator whose matrix 
elements between deformed states
are equal to the overlaps of two states with quasi-particles located at opposite edges:
\begin{multline}
\langle N,M,\underbar{t}\,|{\cal A}_{\rm qp}(\xi)|N,M',\underbar{t}'\,\rangle \equiv
 \int \prod_id^2z_i\\ \times 
\langle N,M,\underbar{t}\,|
\psi_{\rm qp}^\dag(\xi')|\underbar{z}\,\rangle \langle \underbar{z}\,|\psi_{\rm qp}(\xi)
|N,M',\underbar{t}'\,\rangle.
\label{tun-matr}
\end{multline}
Here we face the problem that a wave function of a QH system describing a single quasi-particle,
$\langle \underbar{z}\,|\psi_{\rm qp}(\xi)|N,M,\underbar{t}\,\rangle$, is not well defined.
This is because one is not able to remove a charge $1/m$ form a system consisting of electrons.
Therefore, we first construct the electron tunneling operator ${\cal A}_{\rm el}$ and then 
derive the  quasi-particle tunneling operator from the observation 
that tunneling of $m$ quasi-particles at one point is equivalent to tunneling of an
electron at that point. 

The electron operator is formally defined by
$\langle z_1,\ldots,z_N|\psi_{\rm el}(\xi)|\Psi\rangle =
\sqrt{N+1}\langle z_1,\ldots,z_N,\xi|\Psi\rangle$, which leads to the result:
\begin{multline}
\langle \underbar{z}\,|\psi_{\rm el}(\xi)|N+1,M,\underbar{t}\,\rangle =
\\ \xi^Me^{-|\xi|^2/4l_B^2+m\omega(\xi)}\prod_i(\xi-z_i)^m\langle \underbar{z}\,
|N,M,\underbar{t}\,\rangle,
\label{el-def}
\end{multline}
where we have omitted a combinatorial factor, because it can be absorbed into the tunneling
amplitudes. Then, using expression (\ref{tun-matr}), with operators $\psi_{\rm qp}$
replaced by the operator $\psi_{\rm el}$ from Eq.\ (\ref{el-def}), we obtain
\begin{multline}
\langle N,M,\underbar{t}\,|{\cal A}_{\rm el}(\xi)|N,M',\underbar{t}'\,\rangle
=e^{m\omega(\xi)+m\omega^*(\xi'^*)}\int \prod_id^2z_i \\ \times
(\xi-z_i)^{m}(\xi'^*-z_i^*)^{m}
\langle N,M,\underbar{t}\,|\underbar{z}\,\rangle \langle \underbar{z}\,|N,M',\underbar{t}'\,\rangle,
\label{el-tun}
\end{multline}
where we have dropped the prefactor $(\xi'\xi^*)^{M}\exp[-(|\xi|^2+|\xi'|^2)/4l_B^2)$,
because, for a {\em short} tunneling path, $\xi'\to\xi$, it has no essential
physical meaning and may be
absorbed into the tunneling amplitude. 
It then becomes obvious that the matrix
element of the quasi-particle tunneling operator may be written as
\begin{multline}
\langle N,M,\underbar{t}\,|{\cal A}_{\rm qp}(\xi)|N,M',\underbar{t}'\,\rangle
=e^{\omega(\xi)+\omega^*(\xi'^*)}\int \prod_id^2z_i \\ \times
(\xi-z_i)(\xi'^*-z_i^*)
\langle N,M,\underbar{t}\,|\underbar{z}\,\rangle \langle \underbar{z}\,|N,M',\underbar{t}'\,\rangle,
\label{qp-tun}
\end{multline}
i.e., roughly speaking, we set ${\cal A}_{\rm qp}={\cal A}_{\rm el}^{1/m}$. We stress that this quasi-particle
tunneling operator is unique and well defined. First of all, in terms of the electron coordinates $z_i$, the 
matrix element (\ref{qp-tun}) is a single-valued function, i.e., the criterion (iii) formulated at the beginning
of the Sec.\ \ref{bulk-edge} is satisfied. Second, this matrix element is also a single-valued function
of the coordinates of the tunneling points $\xi$ and $\xi'$. Thus, at the microscopic level, no ambiguity
arises in the definition of a tunneling operators.

In order to evaluate the matrix elements (\ref{qp-tun}),
we first assume that $\underbar{t}'=\underbar{t}$, as in the previous section,
and then generalize our findings.
The product $\prod_i(\xi-z_i)$ in Eq.\ (\ref{qp-tun}) can be rewritten as
$\exp\left[\sum_i\ln(\xi-z_i)\right]$, and one obtains a similar expression
for $\prod_i(\xi'^*-z_i^*)$. Expanding the logarithms in power series (\ref{exp-n})
on the inner, $|\xi|<|z_i|$, and outer, $|\xi'|>|z_i|$, edges, we arrive at the following expression
for the matrix elements (\ref{qp-tun}):
\begin{widetext}
\begin{multline}
\langle N,M,\underbar{t}\,|{\cal A}_{\rm qp}(\xi)|N,M',\underbar{t}\,\rangle
=\exp[\omega(\xi)+\omega^*(\xi'^*)+N\ln\xi'^*]
\\ \times\int \prod_id^2z_i z_i\exp\left\{
-\sum_{k>0}\left[ \frac{\xi^k}{kz_i^k}+\frac{(z_i^*)^k}{k(\xi'^*)^k} \right]\right\}
\langle N,M,\underbar{t}\,|\underbar{z}\,\rangle \langle \underbar{z}\,|N,M',\underbar{t}\,\rangle.
\label{hren}
\end{multline}
Taking into account that
$m \sum_i(z_i)^{-k} \langle \underbar{z}\,|N,M',\underbar{t}\,\rangle =
\partial/\partial t_{-k}\langle
\underbar{z}\,|N,M',\underbar{t}\,\rangle$, and
$m\sum_i (z_i^*)^k \langle
N,M,\underbar{t}\,|\underbar{z}\,\rangle = \partial/\partial
t_k^*\langle N,M,\underbar{t}\,|\underbar{z}\,\rangle$,
we pull the power series in $z_i$ out of the integral. Then we use the
fact that $\prod_i z_i\langle\underbar{z}\,|N,M',\underbar{t}\,\rangle
= \langle\underbar{z}\,|N,M'+1,\underbar{t}\,\rangle$ to
rewrite Eq.\ (\ref{hren}) in the following form:
\begin{multline}
\langle N,M,\underbar{t}\,|{\cal A}_{\rm qp}(\xi)|N,M',\underbar{t}\,\rangle =
\exp[\omega(\xi)+\omega^*(\xi'^*)+N\ln\xi'^*]
\\ \times\exp\left\{-\frac{1}{m}\sum_{k>0}\left[\frac{\xi^k}{k}\frac{\partial}{\partial
t_{-k}}+\frac{1}{k(\xi'^*)^k}\frac{\partial}{\partial
t_k^*}\right]\right\}\langle N,M,\underbar{t}\,|N,M'+1,\underbar{t}\,\rangle.
\label{hren-2}
\end{multline}
\end{widetext}
One can immediately see that the matrix element (\ref{hren-2}) vanishes
unless $M=M'+1$.

At this point, we need to replace $\underbar{t}$ with $\underbar{t}'$ in Eq.\ (\ref{hren-2})
in order to evaluate the matrix element 
$\langle N,M,\underbar{t}\,|{\cal A}_{\rm qp}(\xi)|N,M',\underbar{t}\,\rangle$. 
The best way to proceed is to apply the shifts generated by the derivatives in 
the right hand side of Eq.\ (\ref{hren-2}) directly to the matrix element
$\langle M,N,\underbar{t}\,|N,M,\underbar{t}'\,\rangle$. We, 
then, use Eqs.\ (\ref{free-en}) and (\ref{products}), and apply the definitions (\ref{ak}) together with the
matrix element (\ref{matrixelements})
to arrive at the following expression for the restriction of the tunneling operator (\ref{tun-matr}):
\begin{equation}
{\cal A}_0(\xi) = \exp\big[i\phi_M + N\ln\xi'^* + i\varphi_D(\xi) - i\varphi^\dag_U(\xi')\big],
\label{h-tunnel}
\end{equation}
where we have introduced the index $0$ to indicate that we consider a quasi-particle tunneling operator
for a {\em symmetric} Corbino disk, i.e., before deformations of the QH liquid are taken into account. 
Such deformations are treated in Sec.\  \ref{deformations-gs}.
In (\ref{h-tunnel}), the following fields appear:
\begin{subequations}
\begin{equation}
\varphi_U(\xi') = \sum_{k>0}\frac{1}{\sqrt{k}}\big[(\xi'/r_U)^ka_{kU}+(r_U/\xi')^ka_{kU}^\dag\big],
\end{equation}
\begin{equation}
\varphi_D(\xi) = \sum_{k>0}\frac{1}{\sqrt{k}}\big[(r_D/\xi)^ka_{kD}+(\xi/r_D)^ka_{kD}^\dag \big].
\end{equation}
\label{bsefield}
\end{subequations}

Several remarks are in order. 
One easily sees that the quasi-particle tunneling operator (\ref{h-tunnel}), taking into account 
definitions (\ref{bsefield}), essentially coincides with the tunneling operator found in the effective theory, 
as given by  Eqs.\ (\ref{tun-ef-A}) and (\ref{ex-modes}). These operators differ only in the parametrization of the
boundary $\partial D$ of a QH system [2D coordinates $\xi= r e^{ i \theta }$ in (\ref{h-tunnel}), 
versus 1D coordinates $x = r \theta$ in
(\ref{tun-ef-A})], and by zero modes, which are different in the case of a single edge considered in 
Sec.\ \ref{boundary}.
Furthermore, note that the wave number $k$ in Eqs.\ (\ref{bsefield}) has to be 
replaced by the wave vector: $k\to kr_s$. 
We also mention that
tunneling operators, as defined in (\ref{h-tunnel}), evaluated at two different points commute:
\begin{equation}
[{\cal A}_0(\xi_L),{\cal A}_0(\xi_R)]=0.
\label{commutativity}
\end{equation} 
This is because the two fields $\varphi_U$ and $\varphi_D$ in (\ref{h-tunnel}) have opposite chiralities, and, in the
case of a symmetric Corbino disk considered so far, their contributions to the  
commutator cancel exactly. Is this property a consequence of the symmetry of the
Corbino disk, or can it be extended to an arbitrary geometry? We propose the following argument,
suggesting the universality of the relation (\ref{commutativity}).
In order to arrive at expression (\ref{h-tunnel}) for the quasi-particle
tunneling operator, we have dropped the prefactor $(\xi'\xi^*)^{M}$ in the electron tunneling 
operator (\ref{el-tun}), and, consequently, the prefactor $(\xi'\xi^*)^{M/m}$. This is motivated
by the proximity of the tunneling points in a real experimental situation. If we relax this requirement
and consider arbitrary locations of tunneling points at the boundaries of the Corbino disk
then the commutator $[{\cal A}_0(\xi_L),{\cal A}_0(\xi_R)]$ acquires additional contributions
from zero modes and from the oscillators. It turns out that these contributions cancel exactly,
suggesting the universality of the commutation relation (\ref{commutativity}). 

Next, we find the restricted charge density
operators at the edges of a QH system,
$\rho_D(\theta) = P\sigma_D(\theta)P - M/2\pi m r_D$ and
$\rho_U(\theta) = P\sigma_U(\theta)P + (M + m N)/2\pi m r_U$,
by rewriting the result (\ref{sol-def})
in terms of the operators (\ref{ak}):
\begin{subequations}
\begin{eqnarray}
\rho_D(\theta) = -\frac{1}{2\pi r_D}\partial_\theta\varphi_D(\xi)-\frac{M}{2\pi m r_D}, \\
\rho_U(\theta) = \frac{1}{2\pi r_U}\partial_\theta\varphi_U(\xi)+\frac{M+ m N}{2\pi m r_U}.
\label{rho-edge-mic2}
\end{eqnarray}
\label{rho-edge-mic}
\end{subequations}
Here the homogeneous contributions describe the charge accumulation caused by the variation
of the quantum numbers $M$ and $N$. We note that, again, these results agree with those of the effective theory
presented in Sec.\ \ref{bulk-edge}. In particular, setting $M=0$ in Eq.\ (\ref{rho-edge-mic2})
and changing the parametrization of the boundary, we arrive at the expression (\ref{rho-edge}), complemented
by (\ref{ex-modes}). Also note that Eqs.\ (\ref{rho-edge-mic}) lead to the following
commutation relations
\begin{equation}
\![{\cal A}_0(\xi),\rho_s(\theta)] =
\pm \frac{1}{mr_s} \delta(\theta - \theta_s){\cal A}_0(\xi),
\end{equation}
$s = U,D$, where the angles $\theta_D$ and $\theta_U$ parametrize the position
of the tunneling points in coordinates of the inner and outer edge.
These commutation relations show that the tunneling operator (\ref{h-tunnel})
creates a pair of point-like charges of magnitude $\pm 1/m$.

Finally, we must find the restriction of the operators of electron tunneling from the quantum Hall
edges to the Ohmic contacts. This can be done by applying the  technique used above
to the electron annihilation operator (\ref{el-def}). The result of the restriction
is given by
\begin{multline}
{\cal A}_{U} = c_U^\dag \exp\big[im\varphi_U(\xi_U)\big] \\ \times\exp\big[-i\phi_N + (mN+M)\ln\xi_U\big]
\label{elec-u}
\end{multline}
for tunneling from the outer edge to the upper Ohmic contact (see Fig.\ \ref{mz-scr} for notations),
while the tunneling operator on the inner edge is given by
\begin{multline}
{\cal A}_{D} = c_D^\dag \exp\big[im\varphi_D(\xi_D)\big] \\ \times \exp\big[-i\phi_N + im\phi_M +  M\ln\xi_D\big],
\label{elec-d}
\end{multline}
where $c_U$ and $c_D$ are electron annihilation operators in Ohmic reservoirs.

\subsection{Deformations of the ground state }
\label{deformations-gs}

We are now in a position to investigate the effects of a deformation of the symmetric Corbino disk
upon the collective and local excitations. For this purpose, we consider the combined 
effect of a ground state deformation, described by some function $\omega_d(z)=\sum_k t_{dk}z^k$, and of
deformations caused by excitations, $\omega(z)=\sum_k t_kz^k$. The total deformation function is then given by
\begin{equation}
\omega_{\rm tot}(z)=\omega(z)+\omega_{d}(z)=\sum_k (t_k+t_{dk})z^k.
\label{tot-def}
\end{equation}  
This function has to be plugged into the wave function (\ref{wave-1}). According to the classical plasma 
analogy, the deformation $\omega_d$ contributes to the charge density at the edge, as given in Eqs.\ (\ref{sol-def}). 
However, since we consider the deformation of a ground state, e.g., by a modulation gate, the Coulomb energy
(\ref{edge-spect}) of the overall deformation does not change. We therefore choose to maintain the definition
(\ref{ak}), so that the Hamiltonian (\ref{h-edge}) of the collective modes remains unchanged.

The ground state deformation also affects the adjoint operators of the collective modes.
This follows by considering the scalar products of the deformed wave functions:
\begin{multline}
\langle N, M,\underbar{t}\,| N, M,\underbar{t}'\,\rangle
\\ =\exp \big[-mF_{\rm pl}(t^*_k+t^*_{dk}, t'_k+t_{dk})\big],
\label{free-gate}
\end{multline}
where $F_{\rm pl}$ is given by Eq.\ (\ref{products}). Repeating the steps that lead to Eqs.\ (\ref{a-ku})
and (\ref{a-kd}), we find that the adjoint operators acquire a shift:
\begin{subequations}
\begin{eqnarray}
a^\dagger_{kU}\to a^\dagger_{kU}+i\sqrt{k}(r_U^kt^*_{dk}+r_U^{-k}t_{d,-k}),\\
a^\dagger_{kD}\to a^\dagger_{kD}+i\sqrt{k}(r_D^kt_{dk}+r_D^{-k}t^*_{d,-k}).
\end{eqnarray}
\label{shifts}
\end{subequations}
Obviously, this shift does not change the canonical commutation relations (\ref{commut-mic}).
 
Next, we construct the modified quasi-particle tunneling operator ${\cal A}_{\rm qp}(\xi)$. For this purpose,
we repeat the steps starting with Eq.\ (\ref{qp-tun}) and leading to the result (\ref{h-tunnel}), but
using new collective mode operators (\ref{shifts}) and the ``deformation function'' (\ref{tot-def}).
The result of the calculations is the expression
\begin{equation}
{\cal A}_{\rm qp}(\xi) = {\cal A}_0(\xi)\exp\left\{2i\,{\rm Im}[\omega_d^-(\xi)-\omega_d^+(\xi')]\right\},
\label{new-tun-def}
\end{equation}
where we have introduced the functions
\begin{equation}
\omega_d^+(z)=\sum_{k>0}t_{dk}z^k,\quad \omega_d^-(z)=\sum_{k>0}t_{d,-k}z^{-k},
\label{parts}
\end{equation}
which are the parts of $\omega_d=\omega_d^++\omega_d^-$, that are holomorphic 
inside the inner edge and outside of the outer edge of the Corbino disk, respectively.
Thus, the only effect of deformations of the ground state of the symmetric Corbino disk
is the appearance of the phase shift in the quasi-particle tunneling operator. The structure
of this term is quite natural: deformations of the outer edge lead to a phase shift
in the field $\varphi_U$ via the function $\omega_d^+$, while deformations of the inner
edge, controlled by the function $\omega_d^-$, shift the field $\varphi_D$.

Next, we note that using Eqs.\ (\ref{sol-def}) one may express the phase shifts in (\ref{new-tun-def})
in terms of the 1D charge densities accumulated at the edges as a result of the deformations:
$2\partial_\theta {\rm Im}\omega_d^+(\xi')=r_U\sigma_U(\theta)$ and $2\partial_\theta {\rm Im}\omega_d^-(\xi)=-r_D\sigma_U(\theta)$. 
Thus, we conclude that neutral deformations
caused, e.g., by a modulation gate, lead to phase shifts in the tunneling operator that
may be evaluated as simple contour integrals of the charge densities accumulated at the edges. 
In the next section we will consider variations of zero modes, 
$\delta N$ and $\delta M$, leading to specific 
variations of the charge densities at the edge. These effects may be described by the phase shifts
already accounted for in the bare tunneling operator (\ref{h-tunnel}), and the phase 
shifts due to neutral deformations considered above. Taking into account Eqs.\ (\ref{rho-edge-mic}),
we finally arrive at the following simple and general result:
\begin{subequations}
\begin{eqnarray}
{\cal A}_{\rm qp}(\xi) = {\cal A}_0(\xi)\exp\left\{-i\delta\varphi_U(x')+i\delta\varphi_D(x)\right\},\\
\delta\varphi_U=2\pi\!\!\int\! dx'\delta\rho_U(x'),\quad \delta\varphi_D=2\pi\!\!\int\! dx\delta\rho_D(x),
\label{final-tun} 
\end{eqnarray}
\label{genres}
\end{subequations}
where, for convenience, we choose the 1D parametrization, $x'=r_U\theta$ and $x=-r_D\theta$,
which accounts for the opposite chiralities of the edge states, and we recall that ${\cal A}_0$ is the bare
quasi-particle tunneling operator for a symmetric Corbino disk given in Eq.\ (\ref{h-tunnel}). 
The constants of integration in Eqs.\ (\ref{final-tun}), if needed, may be found by evaluating 
functions $\omega_d^+$ and $\omega_d^+$ directly. These are single-valued functions:  
The closed-contour integrals vanish, because the deformations are neutral.
The operators $\delta\varphi_s$ in this expression account for the specific phase shifts 
analyzed in the next section. The operator  ${\cal A}_0$ determines the scaling
behavior of tunneling rates.

To summarize the main results of this section, equations (\ref{h-tunnel}-\ref{elec-d}) 
complete the restriction procedure, and the  resulting low-energy theory of a QH system 
confined to a Corbino disk agrees with the effective theory of Refs.\  [\onlinecite{Wen}] and  
[\onlinecite{Frol}]. However, in addition to this important conclusion, we obtain
information that cannot be extracted from the effective theory alone.
 Namely, the effective theory
does not specify precisely how zero modes enter the tunneling operators. For example, our direct
microscopic calculations suggest that the quasi-particle tunneling operators (\ref{h-tunnel})
do not contain the zero mode $M$ that counts the number of quasi-particles localized in the
central hole of the Corbino disk. Moreover, we do not find any traces of additional Klein factors 
\cite{Safi} in the tunneling operators.

\section{QH interferometer away from equilibrium}
\label{conseq}

In this section we return to the model of a QH interferometer, see Fig.\ \ref{mz-scr},
and investigate various effects  of inserting a singular magnetic flux, applying a modulation gate
voltage, and coupling to Ohmic contacts. The most important effect is the accumulation of a charge
density, $\delta\rho_s$, at the edges $s = U$ or $D$. It leads a the phase shift in the quasi-particle 
tunneling operators, which may be evaluated using Eq.\ (\ref{final-tun}). Electron tunneling 
operators, see (\ref{elec-u}) and (\ref{elec-d}), also acquire phase shifts. But they cancel in tunneling rates.
In contrast, the phase shifts of quasi-particle tunneling operators add to the AB phase in the oscillating
part of the quasi-particle tunneling rates. The consequences of this observation are twofold. First, the AB effect
may be observed by applying a modulation gate voltage or a singular magnetic flux. We will find
the periods of AB oscillations in both cases. Second, the phase shifts $\delta\varphi_s$, and, consequently, 
the quasi-particle tunneling rates,  depend
on the variations of zero modes, $\delta N$ and $\delta M$. 
We will describe tunneling processes and a stationary quasi-particle current on a long
time scale by solving the master equation for the probability distributions in the space of those zero modes.  

\subsection{Ohmic contacts}
\label{Ohmic-c}

So far, we have considered a QH system confined to a Corbino disk and electrically isolated
from the measurement circuit. Such a system, complemented by two QPCs connecting the inner and outer
edges of the Corbino disk, may nevertheless be considered a QH interferometer. The time-dependent
quasi-particle current in such a system may be investigated by using, e.g., the time domain capacitance 
spectroscopy method.\cite{tdcs} As we will see, there is no reason for the quasi-particle AB effect
not to be observed in such measurements, using an MZ interferometer. However, the Byers-Yang theorem
is formulated for a stationary current, and the naive argument presented in the introduction
considers an MZ interferometer as an open system. Experimentally,\cite{mz1} such a situation is realized
by coupling a QH system to Ohmic contacts.

A correct physical description of coupling Ohmic contacts to QH edge states is a complex 
theoretical problem. We feel that, despite several attempts 
\cite{Averin-ohmic,Kane-Fisher}, this problem has not been fully solved.
In particular, we are not aware of a discussion in the literature of how 
different models of Ohmic contacts can be discriminated (or verified) experimentally. 
The difficulty
of modeling Ohmic contacts is related to the fact that Ohmic reservoirs, typically
realized by heavily doping a part of the semiconductor substrate, represent 
physical systems entirely different from a 2DEG in a QH state. In this situation it is 
best to rely on minimal physically plausible requirements to correctly describe an 
Ohmic contact: (i) An Ohmic contact is a metallic reservoir, that has a large capacitance
and, as a result, suppresses fluctuations of the total charge at the QH edges;
(ii) it has a resistance much smaller than that of the system and 
creates negligibly little noise; 
(iii) an Ohmic contact efficiently equilibrates the edge states.
If these requirements are satisfied in a model it becomes 
experimentally difficult to distinguish such a model of an 
Ohmic contact from an ideal one.

Here we focus on the  first requirement and address the other ones in Sec.\ \ref{QPcurrent}.
An ideal Ohmic contact suppresses charge fluctuations at the edges by absorbing
all the excess charges arising from variations of zero modes and of the magnetic field flux.
Thus, no variations of the charge densities $\delta\rho_s$ arise, except at the points 
$\xi_U=r_Ue^{ix'_U/r_U}$ and $\xi_D=r_De^{-ix_D/r_U}$, where the Ohmic contacts are located, 
and at the position $\xi_G=r_Ue^{ix'_G/r_U}$ of 
the modulation gate; see Fig.\ \ref{mz-scr}. Denoting these charges by $Q_U$, $Q_D$,
and $Q_G$, respectively, we may write:
\begin{subequations}
\label{charges}
\begin{eqnarray}
\delta\rho_U(x')&=&Q_U\delta(x'-x'_U)+Q_G\delta(x'-x'_G),
\label{charges1}\\
\delta\rho_D(x)&=&Q_D\delta(x-x_D).
\label{charges2}
\end{eqnarray}
\end{subequations}
These equations have to be substituted into Eqs.\ (\ref{genres}) which determine the phase shifts
in the quasi-particle tunneling operators. 

Next, we evaluate the excess charges in different situations. If the zero mode $N$ changes 
by $\delta N$, due to an electron tunneling from Ohmic contact to an edge, this leads to an accumulation
of charge at the outer edge of the Corbino disk near the upper Ohmic contact; hence
$Q_U=\delta N$. If the zero mode $M$ changes by $\delta M$, as a result of quasi-particle tunneling
at QPCs, this leads to an accumulation of charge at the upper ohmic contact, $Q_U=\delta M/m$,
and depletes the charge at the inner Ohmic contact, $Q_D=-\delta M/m$. If one applies a modulation gate
voltage in order to deform the (outer) path of the interferometer, and thus changes the total magnetic
flux through it by $\Phi_G$, this leads, according to Eq.\ (\ref{dens-g}), to a neutral deformation
and removes a charge $\Phi_G/m\Phi_0$ at the point $\xi_G$. Therefore, in this case, $Q_G=-\Phi_G/m\Phi_0$.
However, since the homogeneous part of the edge density is screened by the Ohmic contacts, 
the excess charge is accumulated at the upper Ohmic contact, and thus $Q_U=\Phi_G/m\Phi_0$.

Finally, adiabatically threading a singular 
magnetic flux $\Phi$ through the hole in the Corbino disk leads to a homogeneous shift of the edges
described by Eq.\ (\ref{flux-shift}). Again, $Q_U=\Phi/m\Phi_0$ and $Q_D=-\Phi/m\Phi_0$, as a result of perfect
screening by ohmic contacts. Adding all the effects considered here, we obtain
\begin{subequations}
\begin{eqnarray}
Q_U&=&\delta N +\frac{\delta M}{m}+\frac{\Phi+\Phi_G}{m\Phi_0},\\
Q_D&=&-\frac{\delta M}{m}-\frac{\Phi}{m\Phi_0},\quad Q_G=-\frac{\Phi_G}{m\Phi_0}.
\end{eqnarray}
\label{charges-res}
\end{subequations}
Already at this level, we can see how the Byers-Yang argument is saved: Changing the 
magnetic flux $\Phi$ by one flux quantum is compensated by a reduction of the number 
of quasi-particles, $\delta M$, by $1$. 

\begin{center}\begin{figure}[htb]\begin{center}
\epsfxsize=7cm
\epsfbox{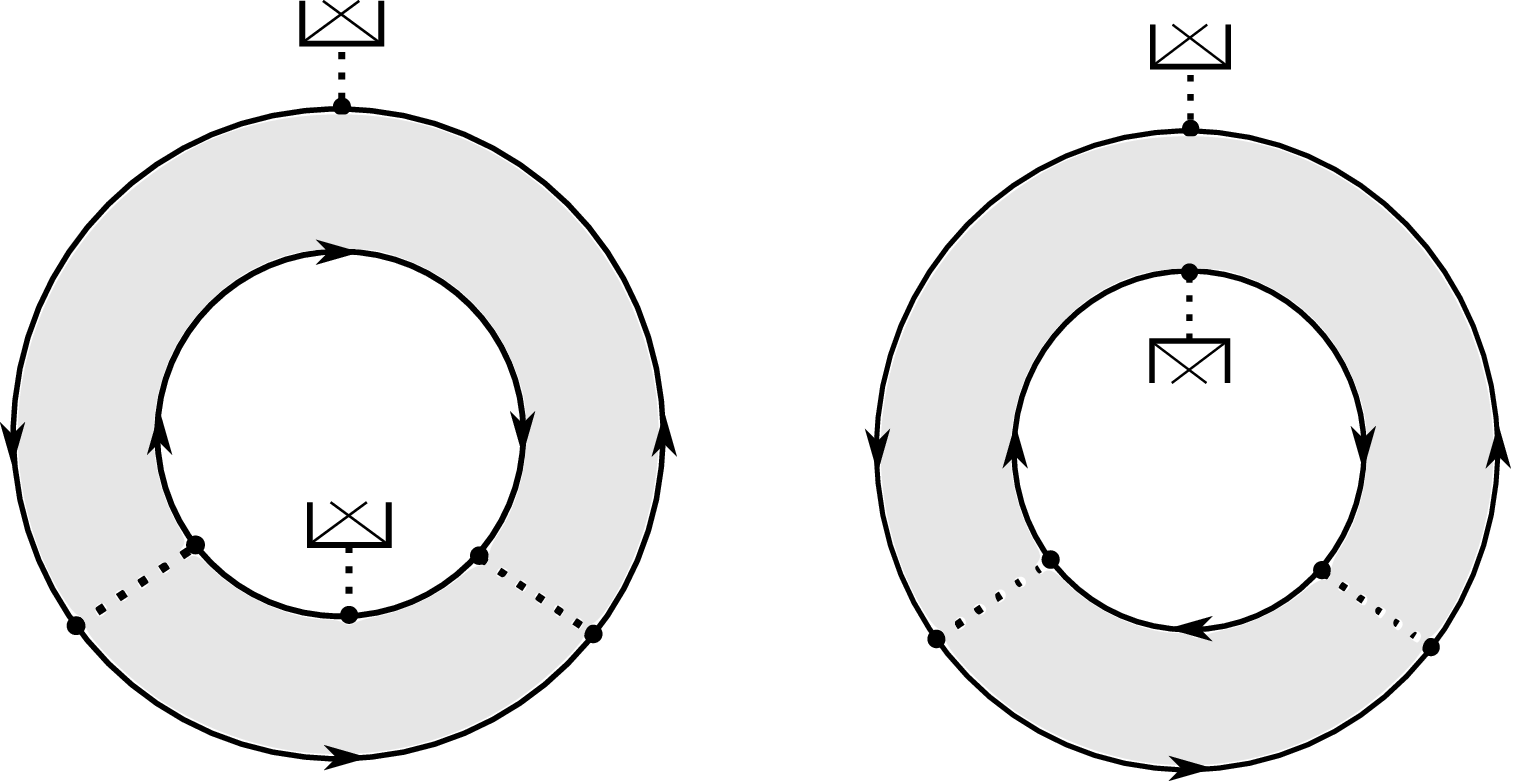}
\caption{Schematic illustration of the interferometers of the MZ type (left panel) and of the FP type
(right panel).}
\vspace{-3mm}
\label{mzfp}
\end{center}\end{figure}\end{center}

The effect of variations of the excess charges is twofold. First, the charges $Q_U$ and $Q_D$
are screened by the Ohmic reservoirs through their own charge capacitances, $C_U$ and $C_D$, which are much
larger than the capacitances of the Corbino disk. Therefore, equation (\ref{ener-zero}) for the 
ground state energy has to be replaced with the following expression
\begin{equation}
E_0=\sum_{s=U,D}\frac{Q_s^2}{2C_s},\qquad \frac{C_sv_s}{r_s}\gg 1,
\label{gs-energy}
\end{equation}
i.e., states corresponding to different values of zero modes are almost degenerate, which 
implies that there are strong fluctuations in equilibrium. We will see, however, that these fluctuations
do not suppress AB oscillations.
Second, as  demonstrated below, to leading order in tunneling processes, the AB oscillation in current 
originate from the term ${\cal A}^\dag_{\rm qp}(\xi_L){\cal A}_{\rm qp}(\xi_R)$, which depends on an integral
of the densities (\ref{charges}) over a closed path and thus depends on excess charges.
However, before evaluating this term, we pause for an important remark.

By appropriately choosing the positions of the Ohmic 
contacts relative to the positions of the QPCs, as 
shown in Fig.\ \ref{mzfp}, one can model both FP and MZ type interferometers with the Corbino disk topology.
If the Ohmic contacts are located at the positions illustrated 
in the left panel, there are only
two coherent paths connecting them, which describes an MZ interferometer. In contrast, if the 
Ohmic contacts are located at the positions shown on the right panel, 
then there are several coherent paths 
connecting them, differing in the number of reflections between the QPCs, as for an FP-type
interferometer. 

\begin{center}\begin{figure}[htb]\begin{center}
\epsfxsize=3.5cm
\epsfbox{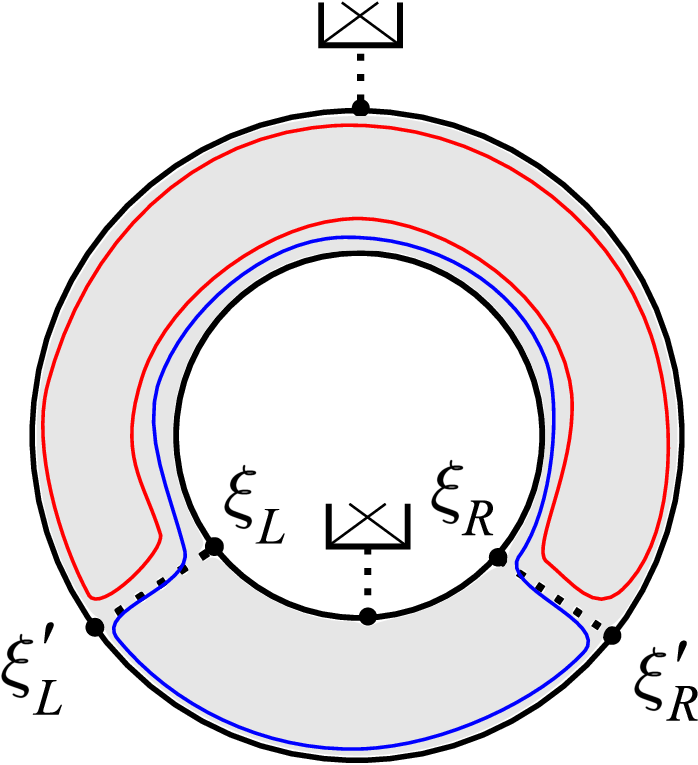}
\caption{Two possible choices of the integration path in Eqs.\ (\ref{MZ-phase}), (\ref{FP-phase}),
and (\ref{eff-phas}) are shown. 
In both cases the result is the same: If one chooses the blue contour, then the contributions 
of the edge densities are zero, while the bulk fields give the phase $2\pi (\delta M + \Phi/\Phi_0)$.
If one chooses the red contour, then the contribution of the bulk fields is zero, since the contour does not 
enclose the hole. However, the contribution of the
charge density, accumulated at the edge near the Ohmic contact, gives the contribution 
exactly equal to the calculated above.}
\vspace{-3mm}
\label{twopath}
\end{center}\end{figure}\end{center}

We expect that the singular magnetic flux $\Phi$ threading the hole of the Corbino disk
enters differently in the expressions of the quasi-particle tunneling operators, for an MZ and an FP interferometer, respectively.
Indeed, when evaluating the term ${\cal A}^\dag_{\rm qp}(\xi_L){\cal A}_{\rm qp}(\xi_R)$ with the help
of expressions (\ref{genres}), (\ref{charges}) and (\ref{charges-res}), one must take into
account the position of the inner Ohmic contact, which leads to
\begin{multline}
{\cal A}^\dag_{\rm qp}(\xi_L){\cal A}_{\rm qp}(\xi_R)={\cal A}^\dag_0(\xi_L){\cal A}_0(\xi_R)\\
\times
\exp\left\{2\pi i\left(\frac{\delta M}{m}+\frac{\Phi+\Phi_G}{m\Phi_0}\right)\right\}
\label{MZ-phase}
\end{multline}
for an MZ-type interferometer, and
\begin{equation}
{\cal A}^\dag_{\rm qp}(\xi_L){\cal A}_{\rm qp}(\xi_R)={\cal A}^\dag_0(\xi_L){\cal A}_0(\xi_R)
\exp\left\{\frac{2\pi i\Phi_G}{m\Phi_0}\right\}
\label{FP-phase}
\end{equation}
for an FP-type interferometer.\cite{footnote-Klein} The difference in these results matches the observation that 
the interference paths in an FP interferometer do not enclose a singular magnetic flux. We also note that, when evaluating
the phase shifts in (\ref{MZ-phase}) and (\ref{FP-phase}) as integrals of the densities $\delta\rho_s$
over closed paths, there are four possibilities to choose these paths. Two of them are shown in
Fig.\ \ref{twopath}: one is connecting the tunneling
points in the lower part of the Corbino disk, and the other encloses the upper part of it. As one can 
see from Eqs.\ (\ref{charges-res}), the difference is equal to $2\pi\delta N$, i.e., the two paths yield identical phase changes.

Finally, we note that all the results obtained here may be formulated at the level of the effective theory. 
One can use Eqs.\ (\ref{rho-edge}) and (\ref{tun-ef}) in order to find the phase factor in  
the product of the two tunneling operators:
\begin{multline}
 {\cal A}_{\rm qp}^\dag(\xi_L){\cal A}_{\rm qp}(\xi_R) \propto \exp\Big\{ 2\pi i\int dx [\delta\rho_D(x) - \delta\rho_U(x)]\Big. \\ 
 \Big.+ i\int_\gamma B_\mu dr^\mu + (2i/m)\int_\gamma A_\mu dr^\mu\Big\},
\label{eff-phas}
\end{multline}
where the contour $\gamma$ connects the tunneling points $\xi_L$, $\xi_L'$, $\xi_R'$ and $\xi_R$, 
and the accumulated edge densities $\delta\rho_s$ 
are integrated along the corresponding sections of the edges. 
It follows from the action (\ref{s-bulk}) that  
$\langle B_\mu\rangle = -A_\mu/m + {\cal B}_\mu$, where the second term is the 
topological contribution of the charge localized at the inner edge, i.e.,   
$\int_\gamma {\cal B}_\mu dx^\mu = 2\pi M/m$, for an arbitrary contour $\gamma$ 
enclosing the hole in the interferometer. 
In other words, the bulk and edge degrees of freedom are not completely decoupled
for a non-trivial sample topology. We should mention that the total phase 
(\ref{eff-phas}) is independent of the choice of the contour $\gamma$ (see Fig.\ \ref{twopath}). 
This is because the tunneling operators are  single-valued and 
this property continues to hold at the level of the effective theory.

\subsection{Current through the interferometer}
\label{QPcurrent}

\begin{center}\begin{figure}[htb]\begin{center}
\epsfxsize=7cm
\epsfbox{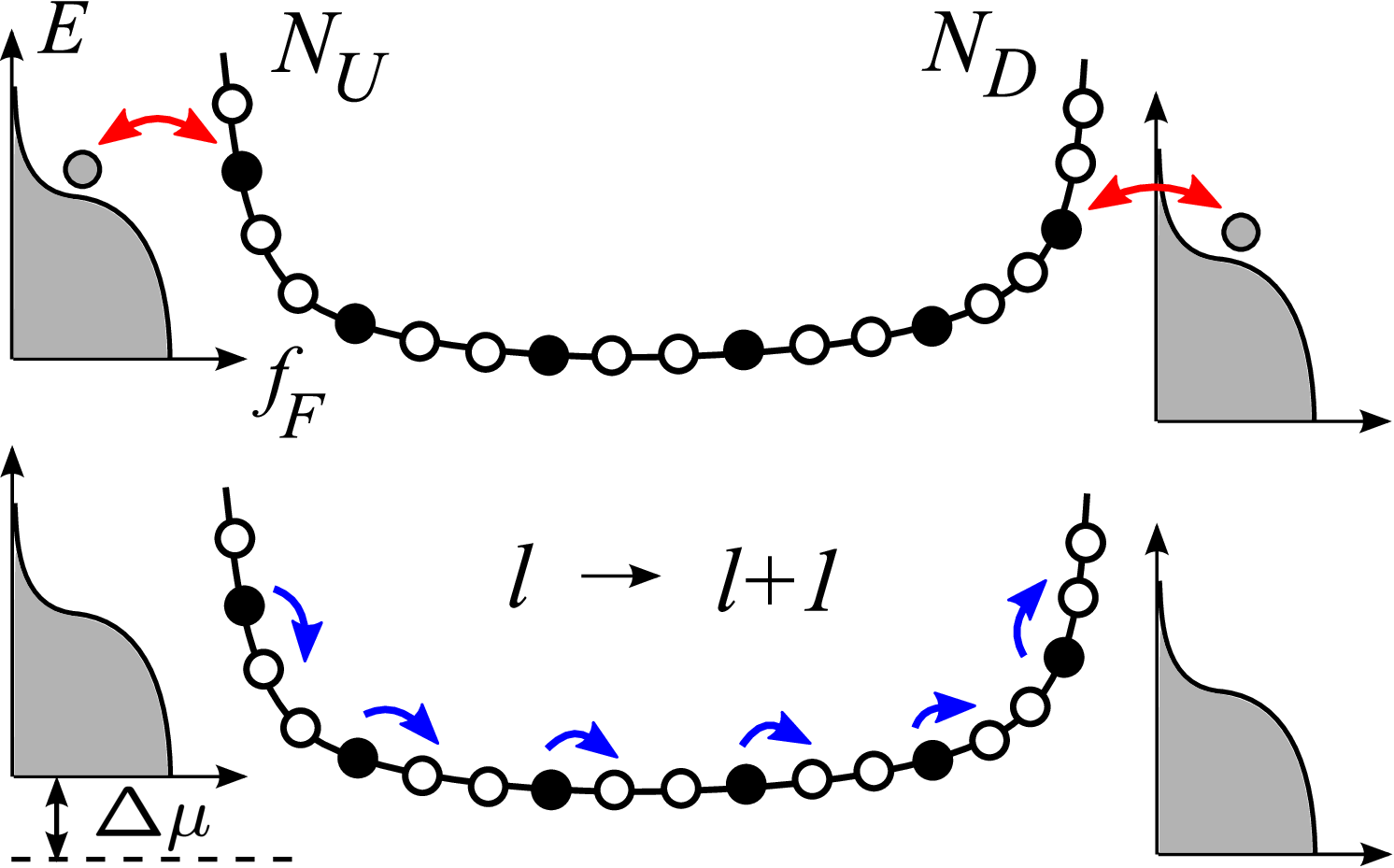}
\caption{Schematic illustration of the processes at Ohmic contacts and QPCs.
{\em Upper panel:} Electrons tunnel between Ohmic contacts and QH edges,
preserving the incompressibility of the QH liquid. These processes change the
numbers $N_U$ and $N_D$. {\em Lower panel:} Tunneling of a quasi-particle
from one edge of the Corbino disk to another leads to the reconstruction
of the wave function and changes the number $l$ by 1. It is accompanied by a change
of the electric charge at the inner edge by the value $1/m$.}
\vspace{-3mm}
\label{relax-scr}
\end{center}\end{figure}\end{center}

To satisfy the second and third requirements on an ideal Ohmic contact (see Sec.\ \ref{Ohmic-c}), 
we consider a QH interferometer with  
strong electron tunneling to Ohmic reservoirs and weak quasi-particle tunneling
between the inner and outer edge (see Fig.\ \ref{relax-scr}). Strong electron coupling to Ohmic contacts
guarantees that the inner and outer edge states are in equilibrium with the metallic reservoirs
with electro-chemical potentials $\mu_U$ and $\mu_D$. The charge current between the Ohmic contacts, which arises as
a response to the potential difference $\Delta\mu=\mu_U-\mu_D$, is due to weak quasi-particle tunneling
at the QPCs. It is convenient to introduce a new notation for zero modes.
We denote by $N_U$ and $N_D$ the numbers of electrons at the outer and inner edge of the interferometer,
respectively.  The number of quasi-particles localized along the inner edge is denoted by $l = 0, \ldots, m-1$:
\begin{subequations}
\begin{eqnarray}
\delta M = -mN_D -l,\\
\delta N = N_U +N_D.
\end{eqnarray}
\end{subequations}
The electron quantum numbers, $N_U$ and $N_D$, change because of tunneling at the Ohmic contacts, while the
quantum number $l$ changes by $1$ when a quasi-particle tunnels from one edge to the other one, as illustrated
in Fig. \ref{relax-scr}.

We consider the zero modes to be classical variables and derive a Master Equation
for the probability distribution functions. Quantum coherence manifests itself in oscillations
of the quasi-particle tunneling rates as functions of the magnetic fluxes $\Phi$ and $\Phi_G$. 
These oscillations originate
from the interference of the two quasi-particle tunneling amplitudes at the left and right QPC. Formally,
these oscillations stem from the $\Phi$-dependent phase factor in the tunneling operators.
Strong coupling to Ohmic contacts implies
that, after every event of quasi-particle tunneling, the edge states relax to the equilibrium state described
by the probability distribution function
\begin{equation}
P_l(N_U,N_D) = \frac{1}{Z_l}\,e^{-\beta(E_0-\sum_s\mu_sN_s)},
\label{rho-eq}
\end{equation}
where $\beta$ is the inverse temperature 
and $Z_l$ is the partition function. The probability $P_l(N_U,N_D)$ depends on the number $l$ of quasi-particles
via the ground-state energy $E_0(N_U,N_D,l)$, which is given in Eqs.\ (\ref{charges-res}) and (\ref{gs-energy}). 
We observe that one does not need to
specify the precise form of coupling to the Ohmic contacts, because the only role of this coupling is to equilibrate the
edge states.  \cite{footnote-ohmic}

To lowest order in quasi-particle tunneling, the distribution function of zero modes may be written as
\begin{equation}
P(N_U,N_D,l) = {\cal P}_lP_l(N_U,N_D),
\label{rho-fact}
\end{equation}
where ${\cal P}_l$ is the probability of finding the system in a state with $l$ quasi-particles, and $P_l(N_U,N_D)$
plays the role of a conditional probability of finding the interferometer in a state with $N_U$ and $N_D$
electrons at the edges, given the number $l$ of quasi-particles.
Considering quasi-particle tunneling as a weak process that changes the number $l$, we may then describe 
the distribution function with the help of a master equation
\begin{equation}
\dot{\cal P}_l = \Omega^+_{l-1}{\cal P}_{l-1}+\Omega^-_{l+1}{\cal P}_{l+1}
-(\Omega^+_{l}+\Omega^-_{l}){\cal P}_{l},
\label{master-eq-qp}
\end{equation}
where $\Omega^+_l$ and $\Omega^-_l$ are the rates of transition from a state with $l$ 
quasi-particles to a state with $l+1$ and $l-1$ quasi-particles, respectively. 
These rates are given by the expression:
\begin{equation}
\Omega^\pm_l = \sum_{N_U,N_D}W^\pm_l(N_U,N_D)P_{l}(N_U,N_D),
\label{t-sum}
\end{equation}
where $W^\pm_l(N_U,N_D)$ are the rates of quasi-particle tunneling between two states 
with a fixed numbers of electrons, $N_U$ and $N_D$.

At time scales much larger than the characteristic tunneling times, the interferometer reaches a steady
state regime, with $\dot{\cal P}_l =0$. It is easy to see that, in this regime, the following quantity is
independent of $l$
\begin{equation}
I=\Omega^+_l{\cal P}_l-\Omega^-_{l+1}{\cal P}_{l+1}.
\label{conserv}
\end{equation}
When $\dot{\cal P}_l =0$ this quantity is actually the charge current that we are looking for. This follows from the expression for
the current
\begin{equation}
I = (1/m)\sum_{l}(\Omega^+_l -\Omega^-_l){\cal P}_l,
\label{curr-d}
\end{equation}
and from the periodic boundary condition, ${\cal P}_0={\cal P}_m$, which can be verified
using Eqs.\ (\ref{rho-eq}) and (\ref{rho-fact}). Note that the detailed balance equation
$\Omega^+_l{\cal P}_l=\Omega^-_{l+1}{\cal P}_{l+1}$ is satisfied
only if $I=0$. Thus, the QH interferometer
in a non-equilibrium steady-state represents an interesting example of a system with broken ``detailed balance''.

We evaluate the tunneling rates $W^\pm_l(N_U,N_D)$ to leading order in the tunneling
Hamiltonian  ${\cal H}_T={\cal A}+{\cal A}^\dagger$,
where ${\cal A}\equiv {\cal A}_{\rm qp}(\xi_L) + {\cal A}_{\rm qp}(\xi_R)$.
A straightforward calculation based on Fermi's Golden Rule yields the following expression:
\begin{multline}
W^+_l(N_U,N_D) = \\ \!\int\! dt\, \mathrm{Tr}\{\rho_{\rm eq}\mathbb{P}(N_U,N_D,l){\cal A}^\dag(t)
{\cal A}(0)\},
\label{t-fix-N}
\end{multline}
and a similar expression for $W^-_l(N_U,N_D)$.
Here the operator $\mathbb{P}(N_U,N_D,l)$ projects onto states with given numbers $N_s$ and $l$,
and the operator $\rho_{\rm eq}$ is the equilibrium density matrix for the oscillators.
We now consider the solution of the master equation (\ref{master-eq-qp}), with tunneling rates 
given by (\ref{t-fix-N}),
for two possible types  of QH interferometers (see Fig.\ \ref{mzfp}).

It turns out that one can find the periodicity of the current without an explicit evaluation of the integral in (\ref{t-fix-N}). 
Indeed, for the MZ interferometer, the product of two tunneling operators, which yields the coherent contribution to (\ref{t-fix-N}),
has the following dependence on the zero mode $l$ and the fluxes [see Eq.\ (\ref{MZ-phase})]:
\begin{equation}
 {\cal A}_{\rm qp}^\dag(\xi_L){\cal A}_{\rm qp}(\xi_R) 
 \propto \exp\left\{ 2\pi i\left(\frac{\Phi+\Phi_G}{m\Phi_0}-\frac{l}{m}\right)\right\}.
 \label{compare}
\end{equation}
Notice that the magnetic flux enters these products, and, hence, the transition rates (\ref{t-sum}), solely 
in the combination 
$(\Phi+\Phi_G)/\Phi_0-l$. A shift of the flux $\Phi$ by one flux quantum $\Phi_0$ is then compensated by the 
shift $l\to l+1$. The quasi-particle current is given by Eq.\ (\ref{curr-d}), where the
probabilities satisfy Eq.\ (\ref{master-eq-qp}), with $\dot{\cal P}_l =0$, with the constraint that $\sum_l{\cal P}_l=1$. All these equations
are periodic in $l$ with period equal to $m$ and invariant under the replacement $\Omega^\pm_l\to \Omega^\pm_{l+1}$.
This implies that the average current has the electronic periodicity, $I(\Delta\mu, \Phi) = I(\Delta\mu, \Phi+\Phi_0)$,
{\it in agreement with the Byers-Yang theorem}. 

For example, the solution of Eqs.\ (\ref{conserv}) and (\ref{curr-d}), for $\nu=1/3$, predicts an average current
\begin{equation}
I=\frac{\Omega^+_2\Omega^+_1\Omega^+_0-\Omega^-_2\Omega^-_1\Omega^-_0}{\sum_l(\Omega^+_{l+1}\Omega^+_l
+\Omega^+_{l+1}\Omega^-_l+\Omega^-_{l+1}\Omega^-_l)}\,,
\label{qp-current}
\end{equation}
which is an explicitly periodic function of $\Phi$ with the electronic period $\Phi_0$.
We note that each tunneling rate $\Omega^\pm_l$ has, however, a quasi-particle periodicity. 
Therefore the quasi-particle periodicity with respect to a singular flux may in principle be observed via, 
e.g., current noise measurements at finite frequencies. This last fact does not contradict the Byers-Yang theorem, 
because this theorem applies only to a stationary state and to long-time measurements. 

In contrast to our findings for the MZ interferometers, it follows from Eq.\ (\ref{MZ-phase}) that the coherent combination 
of the tunneling amplitudes does {\it not} depend on $l$ for FP interferometers.
The current has therefore a quasi-particle periodicity, with period $m\Phi_0$, with respect 
to a modulation gate. We emphasize that this does not violate the Byers-Yang theorem. 
The quasi-particle periodicity with respect to a modulation gate allows one to perform edge spectroscopy, 
as proposed in Ref.\ [\onlinecite{our-frac}].
The product (\ref{FP-phase}) does however {\em not} depend on the singular flux, however, which is natural, 
because the interference contour does not 
wind around the flux tube. The periodicity of the current in the singular magnetic flux is thus 
determined entirely by the Coulomb blockade effect, if present in case of weak coupling to Ohmic contacts,
and is equal to $\Phi_0$.

\section{Conclusion}

Recently, the physics of AB oscillations in electronic interferometers has been the subject of some debate.
A number of authors have claimed that only electronic periodicity may be observed in transport phenomena in Mach-Zehnder interferometers
built from QH states at fractional filling factors $\nu = 1/m$. We have briefly reviewed those results in the introduction. 
Here we repeat that the main argument against the observability of AB oscillations with
{\it quasi-particle periods} is based on the Byers-Yang theorem, which states that the steady-state current 
through the interferometer oscillates with a period given by the electronic period $\Phi_0$ when expressed as a function of the singular 
magnetic flux $\Phi$ threading the interferometer's loop.

We have constructed a microscopic model of a QH interferometer to analyze how and in which cases the electron periodicity 
is restored. Our starting point has been the Laughlin wave-function, which is a 
good approximation for the true ground state of a QH liquid at filling factors $\nu = 1/m$.
We have studied small, incompressible deformations of the ground state of an electronic interferometer and restrict the microscopic Hamiltonian 
to the subspace of states describing such deformations. The restricted Hamiltonian
(\ref{h-edge}) and restricted tunneling operators (\ref{h-tunnel}) are consistent with the effective low-energy theory \cite{Wen,Frol}, 
modified in order to take into account the non-trivial topology of the system.
Interestingly, we have not found any traces of additional Klein factors \cite{Safi} in the quasi-particle
tunneling operators. Moreover, these operators
are single-valued in the positions of the tunneling points.  

Crucial ingredients of our model are the Ohmic contacts. We describe them as regions of the edges that
accommodate any excess charge, thus perfectly screening slow charge fluctuations at the edges. Ohmic contacts
have a large capacitance and enhance the equilibration and dephasing of edge states.
The current through the interferometer, caused by a bias voltage applied to the Ohmic contacts, has been determined with the help
of a master equation describing tunneling at QPCs. We have found that the tunneling 
rates in an MZ interferometer oscillate in the number of quasi-particles $l$. 
Therefore, after a resummation over $l$, the current through the interferometer oscillates 
as a function of the magnetic flux
with the electronic period $\Phi_0$. In contrast, for an {\em FP interferometer}, the tunneling 
rates are independent of $l$, and the current oscillates with the period $m\Phi_0$ as a 
function of the magnetic flux controlled with a modulation gate. The Byers-Yang paradox has been resolved.

In conclusion, we note that the microscopic derivation of the effective low-energy theory 
of an electronic interferometer presented in this paper can be 
generalized to states with other filling factors; in particular, to states 
with non-Abelian quasi-particle statistics. Moreover, our results can easily be generalized to systems 
with a different geometry. Although, generally speaking, the physics 
will remain the same, detailed considerations may reveal new interesting results. In view 
of our findings concerning the nature of quasi-particle interference, it would be very interesting 
to reconsider effects of quasi-particle exchange and statistics. 

\begin{acknowledgments}
We thank C.\ W.\ J.\ Beenakker, V.\ Cheianov, and A.\ Korolyuk for valuable discussions.
This work has been supported by the Swiss National Foundation.
\end{acknowledgments}

\appendix

\section{Quantization of Chern-Simons theory}
\label{app-quant}

The purpose of this appendix is to recall some basic aspects of the quantization of 
the topological Chern-Simons theory,\cite{footnote-app} as described by the action
(\ref{bulk-act}), which is invariant under the gauge transformations 
(\ref{transf-app}).
The most important property of the action (\ref{bulk-act}) is that it is ``topological''.
It does not depend on the metric tensor $g_{\mu\nu}$,
and it is invariant under arbitrary smooth transformations of coordinates. 
Indeed, under a change of coordinates,
 $x^\mu \to y^{\mu'} (x^\mu)$, the field $B_\mu$ transforms as $B_{\mu'}\partial y^{\mu'}/\partial x^{\mu} $, 
so that, in the new coordinates, the action takes the following form:
\begin{multline}
S_0[B]=\alpha_0\int d^3x \epsilon^{\mu\nu\lambda}B_\mu\partial_\nu B_\lambda  \\ =
\alpha_0\int \frac{d^3y}{|\det J|} \epsilon^{\mu\nu\lambda} J^{\mu'}_\mu
J^{\nu'}_\nu J^{\lambda'}_\lambda B_{\mu'}\partial_{\nu'} B_{\lambda'},
\end{multline}
where $J^{\mu'}_\mu = \partial y^{\mu'}/\partial x^{\mu}$ is the Jacobian of the transformation.
Next, using the properties of the antisymmetric tensor $\epsilon^{\mu\nu\lambda}$,
\begin{equation}
 \epsilon^{\mu\nu\lambda}J^{\mu'}_\mu
J^{\nu'}_\nu J^{\lambda'}_\lambda =  \epsilon^{\mu'\nu'\lambda'}\det J,
\end{equation}
we find that the form of the action remains unchanged.
Note that the situation is different from the Maxwell-type contribution to the action (\ref{full-act}), 
which explicitly depends on the metric $g_{\mu\nu}$:
\begin{equation}
S_1[B] = \alpha_1\int d^3r \sqrt{\det g} F_{\mu\nu}F_{\lambda\rho}g^{\mu\lambda}g^{\nu\rho}, 
\end{equation}
where $F_{\mu\nu} = \partial_\mu B_\nu - \partial_\nu B_\mu$.
However, this contribution can be neglected in the low-energy limit.

\begin{figure}[htb]
\epsfxsize=8cm
\epsfbox{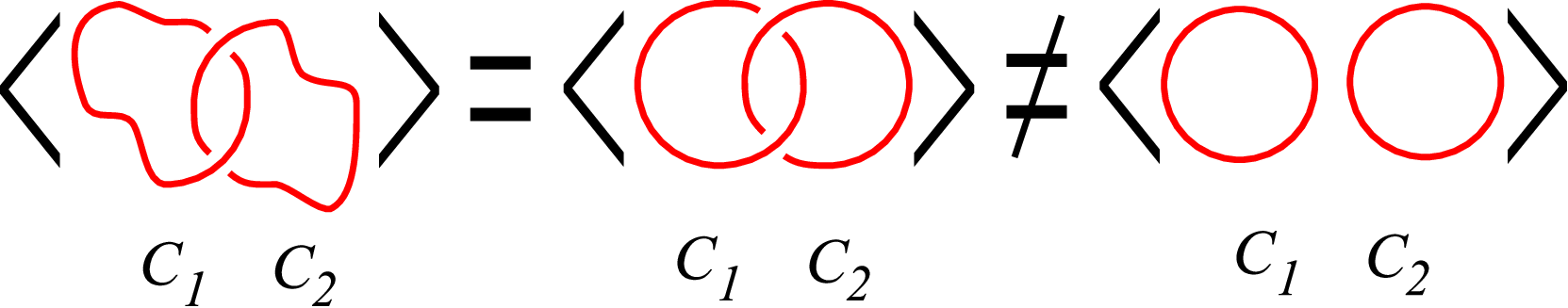}
\caption{Schematic illustration of the topological invariance of the Chern-Simons theory. 
The vacuum average (\ref{link-app}) of the 
flux tube operators corresponding to the contours $C_1$ and $C_2$ (shown by red lines) 
on the left is equal to the average corresponding
the contours shown in the center.
However, the average for the contours shown in the middle differs from the one for the 
contours shown on the right, because the corresponding links 
are topologically inequivalent, i.e., they cannot be transformed into each other by 
smooth deformations.}
\vspace{-3mm}
\label{app-top}
\end{figure}

The topological nature of the action (\ref{bulk-act}) 
implies that the expectation value of an arbitrary observable
depends only on topological properties of this observable. 
For example, the vacuum average of the ``fluxes'' through 
contours $C_1$ and $C_2$  
\begin{equation}
{\cal L}(C_1, C_2) = 2i\alpha_0\langle \oint_{C_1}B_\mu dx^\mu\oint_{C_2}B_\nu dx^\nu\rangle, 
\label{link-app}
\end{equation}
known as the linking number,
does not depend on the shapes
of the contours but only on the way in which $C_1$ and $C_2$ are linked, 
as illustrated in Fig.\ \ref{app-top}. 
This is because the shape of the contours can
be changed with the help of coordinate transformations, but 
the action is invariant under such transformations. 

In the following, we analyze the equations of motion for the field $B_\mu$ 
in order to identify physical states of the Chern-Simons theory. 
Variation of the action (\ref{bulk-act}) with respect to $B_\mu$ yields:
\begin{equation}
\epsilon^{\mu\nu\lambda}\partial_\mu B_\lambda = 0.
\label{eq-mot-app}
\end{equation}
Introducing time and spatial components, $B_\mu = (B_t, \vec{B})$, as well as the 
``electric'' field $\vec{E} = \vec{\nabla} B_t - \partial_t \vec{B}$ and ``magnetic'' 
field $B = \partial_x B_y - \partial_y B_x$, we rewrite Eq.\ (\ref{eq-mot-app}) as follows:
\begin{eqnarray}
 E_x = E_y = B = 0.
\end{eqnarray}
Thus, the dynamics of a QH liquid is similar to the dynamics of a condensate in a 
superconducting metal. 
In a topologically trivial region,
the only solution of the equations of motion (\ref{eq-mot-app}) is 
\begin{equation}
B_\mu = 0,
\end{equation}
up to the gauge transformations (\ref{transf-app}). Physically, 
this behavior reflects the presence of a gap for charge density 
excitations in the bulk, which do not appear in the effective theory.
We conclude that the dynamics of a QH liquid is trivial and the field strength of 
the vector potential $B_\mu$ vanishes, unless some ``topological
defects'' are present. 

\begin{figure}[htb]
\epsfxsize=8cm
\epsfbox{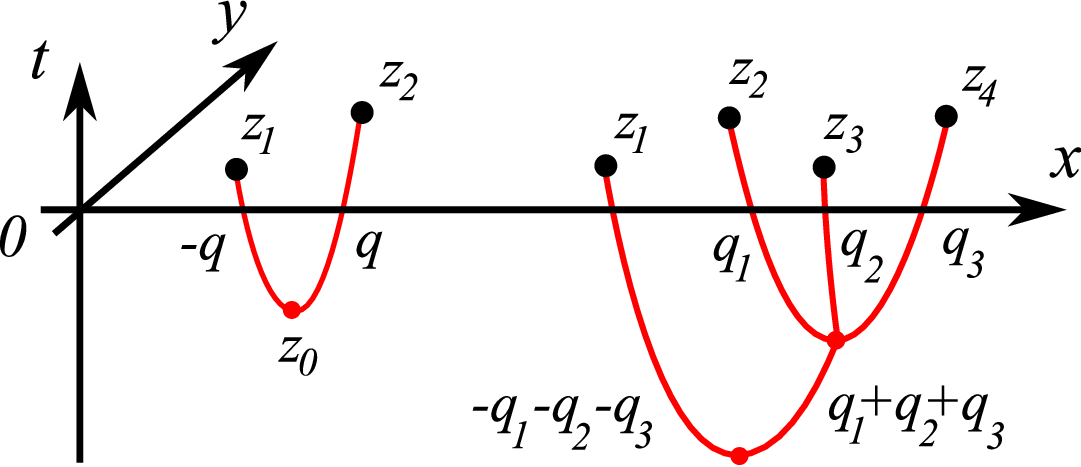}
\caption{Simple examples of states in the Chern-Simons theory,
created by Wilson lines (shown in red color). 
The state shown on the left is created by the operator 
$\exp [iq\int_{z_0}^{z_1}B_\mu dx^\mu -iq\int_{z_0}^{z_2}B_\mu dx^\mu]$. 
All the endpoints (shown by black dots) lie on the plane $t=0$. 
The state shown on the right is a more complicated example of an allowed state. 
Note the ``charge conservation'' condition (\ref{conserv-app}) 
at all nodes (shown by red dots).}
\vspace{-3mm}
\label{app-states}
\end{figure}

The topological defects, which bear some similarities with Abrikosov vortices 
in superconductors, 
may be described by the so called Wilson line operators:
\begin{equation}
 {\cal W}[\underbar{C}\,] = \exp\left\{i\sum_iq_i\int_{C_i}B_\mu dx^\mu\right\}.
\label{wilson-app}
\end{equation}
They are parameterized by a set $\underbar{C} = \{C_i, q_i\}$ 
of contours $C_i$ and real numbers $q_i$.
The operator (\ref{wilson-app}) creates vortices along contours $C_i$ 
with ``vorticities'' $q_i$. 
It turns out that, as we show in the following, these vortexes are 
local quasi-particle excitations with charges proportional to $q_i$. 
The space of states of the QH liquid described by the action (\ref{bulk-act}) 
is then defined as the linear span of the vectors 
\begin{equation}
 |\underbar{C}\,\rangle = {\cal W}[\underbar{C}\,]|0\rangle,
\label{def-states-app}
\end{equation}
corresponding to Wilson lines ending on the surface $t=0$ applied to the ground state $|0\rangle$ 
and satisfying the condition 
\begin{equation}
\sum_i q_i =0 
\label{conserv-app}
\end{equation}
at all nodes of vortices. 
Condition (\ref{conserv-app}) ensures the gauge-invariance of the theory. 
Two examples of Wilson lines are shown in Fig.\ \ref{app-states}.
Lines with the same endpoints $z_i$ and vorticity $q_i$, which are topologically equivalent, 
create the same state.

\begin{figure}[htb]
\epsfxsize=8cm
\epsfbox{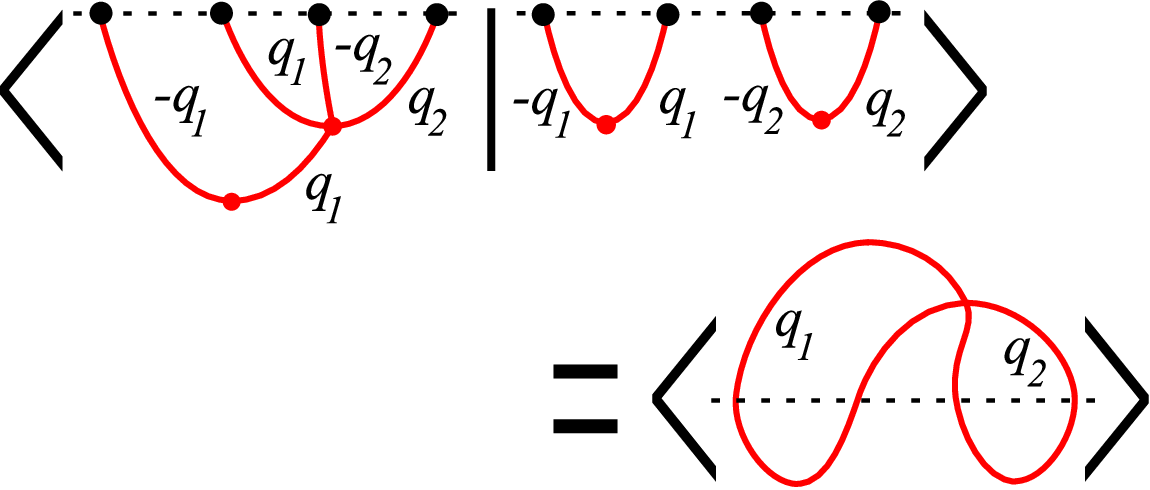}
\caption{An example of the scalar product (\ref{scalar-def-app}).
States on the left and right are described by corresponding Wilson lines. 
The dashed line shows the position of the plane $t = 0$.  
The Wilson line creating the state on the left must be reflected at the plane $t=0$ 
and hooked up to the Wilson line creating the state on the right.
 The scalar product is given by the average of the resulting (closed) Wilson loop, 
which can be evaluated with the help of Eq.\ (\ref{av-app}).}
\vspace{-3mm}
\label{app-scalar}
\end{figure}

The scalar product of two arbitrary states (\ref{def-states-app}) is given by 
the expression  
\begin{equation}
 \langle {\underbar{C}\,}_1|{\underbar{C}\,}_2\rangle = \langle {\cal W}[{\underbar{C}\,}^*_1+{\underbar{C}\,}_2]\rangle,
\label{scalar-def-app}
\end{equation}
where ${\underbar{C}\,}^*_1$ denotes the set of contours,obtained by time reversal
of contours of the set ${\underbar{C}\,}_1$. This procedure is illustrated
in Fig.\ \ref{app-scalar}. Below we show that the scalar products of vectors created by Wilson 
lines with different endpoints 
or with non-matching vorticities $q_i$ vanish, i.e., the corresponding states are orthogonal. 
For non-vanishing products, the average on the right hand side of Eq.\ (\ref{scalar-def-app}) 
can be found by evaluating the Gaussian path integral:
\begin{equation}
 \langle {\cal W}_{\underbar{C}\,} \rangle = \int {\cal D}B_\mu \exp\left\{iS_0[B]+i\sum_iq_i\oint_{C_i}B_\mu dx^\mu\right\}.
\label{av-app}
\end{equation}
The set ${\underbar{C}\,}={\underbar{C}\,}^*_1+{\underbar{C}\,}_2$ consists of only closed contours. 
Thus, taking into account condition (\ref{conserv-app}),
we conclude that all scalar products of states of the theory are gauge-invariant. 
Definition (\ref{scalar-def-app}) of the scalar product completes the construction of the 
{\it quantum} Chern-Simons theory. In what follows, 
we use this definition to identify properties of the excitations created by Wilson lines.

\begin{figure}[htb]
\epsfxsize=8cm
\epsfbox{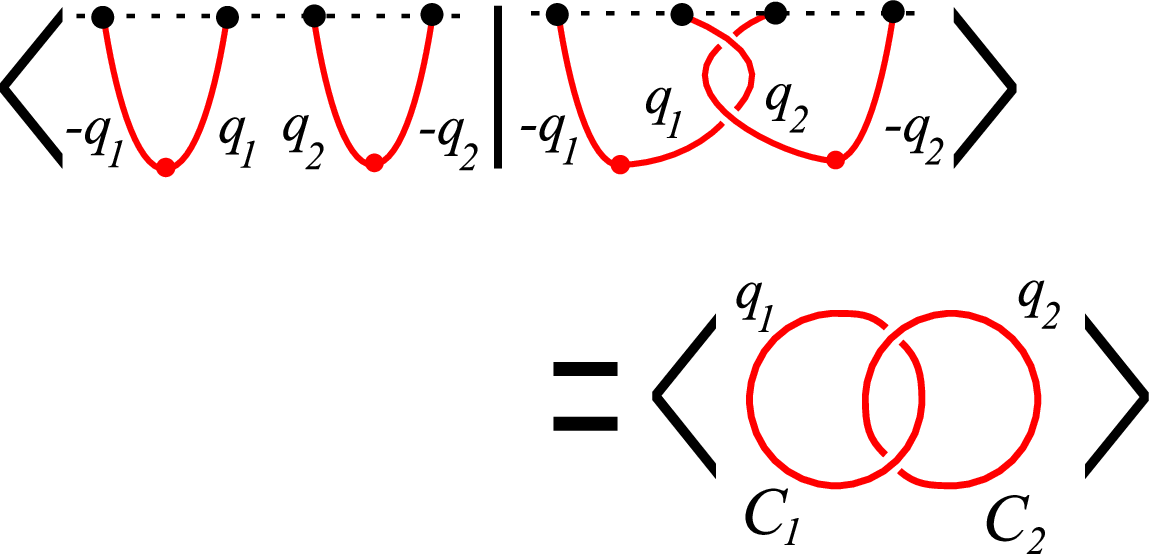}
\caption{Illustration of the calculation of the statistical phase $\theta_{12}$ of two excitations labeled by $q_1$ and $q_2$. 
The scalar product of the two states shown on the left hand side is equal to $\exp(2i\theta_{12})$. According to 
Eqs.\ (\ref{scalar-def-app}) and Eq.\ (\ref{step2-app}), the expectation of the resulting Wilson loop on the right hand side
is proportional to the exponential of the linking number of the two loops $C_1$ and $C_2$. }
\vspace{-3mm}
\label{app-stat}
\end{figure}

One of the most important properties of local excitations is related to their statistical phase. 
In Chern-Simons theory, the statistical phase $\theta_{12}$ of two excitations 
labeled by $q_1$ and $q_2$ is defined as the relative phase of the wave functions 
that differ by braiding of the corresponding Wilson lines. In the example shown in 
Fig.\ \ref{app-stat}, the relative phase of the wave functions is given by $\exp(2i\theta_{12})$. 
In order to find this phase, one has to evaluate the corresponding functional integral. 
Using the Gaussian nature of the functional integral (\ref{av-app}) and the definition 
(\ref{link-app}) of the linking number, we find that 
\begin{equation}
\theta_{12} = \frac{iq_1q_2}{2}\Big\langle \oint_{C_1}B_\mu dx^\mu\oint_{C_2}B_\mu dx^\mu\Big\rangle 
=\frac{q_1q_2}{4\alpha_0}{\cal L}(C_1, C_2).
\end{equation}
The linking number may formally be expressed in terms of the correlation
function $G_{\mu\nu}(x-y) = \langle B_\mu(x) B_\nu(y)\rangle$  
as follows
\begin{equation} 
{\cal L}(C_1, C_2)= 2i\alpha_0\oint_{C_1} dx^\mu \oint_{C_2} dy^\mu G_{\mu\nu}(x-y).
\end{equation}
Since the action (\ref{bulk-act}) is quadratic, $G_{\mu\nu}$
is the Green function of the equations of motion (\ref{eq-mot-app}):
\begin{equation}
\epsilon^{\mu\rho\lambda}\partial_\lambda G_{\rho\nu}(x-y) = ( i/2\alpha_0)\delta^\mu_\nu \,\delta(x-y).
\label{step1-app}
\end{equation}

Using Eq.\ (\ref{step1-app}) and applying Stokes' theorem
to the integral over the contour $C_1$, we find that
\begin{equation}
{\cal L}(C_1, C_2) = \int_{D_1} n_\mu d^2x\oint_{C_2}dy^\mu \delta(x-y),
\label{step2-app}
\end{equation}
where $n_\mu$ is the unit vector field orthogonal to a surface $D_1$ bounded by $C_1$. 
Thus, we find that the linking number of the contours $C_1$ and $C_2$ 
counts how many times one of the contours pinches
through a surface bounded by the other contour. This number is obviously independent of 
the choice of the surface and is topological. 
In the example illustrated on the right hand side of Fig.\ \ref{app-stat}, this number 
equals one, and therefore, the statistical phase is
\begin{equation}
 \theta_{12} = \frac{q_1q_2}{4\alpha_0}.
\label{braid-app}
\end{equation}

Another important property of local excitations is their charge. 
First, we note that the Wilson lines (\ref{wilson-app}) correspond to open contours and are therefore not gauge-invariant. 
The states (\ref{def-states-app}) transform 
under gauge transformations (\ref{transf-app}) as follows:
\begin{equation}
 |\underbar{C}\,\rangle \to \exp\left\{i\sum_iq_i\beta(z_i)\right\}|\underbar{C}\,\rangle.
\end{equation}
In other words, these states transform as wave functions of charge sources localized at the points $z_i$.
It follows then that any two states, $|\underbar{C}\,\rangle$ and $|\underbar{C}'\,\rangle$, with different end points 
of Wilson lines or with non-matching $q_i$ are orthogonal. This is because the scalar product of these states vanishes upon 
integrating over the gauge function $\beta$.

\begin{figure}[htb]
\epsfxsize=8cm
\epsfbox{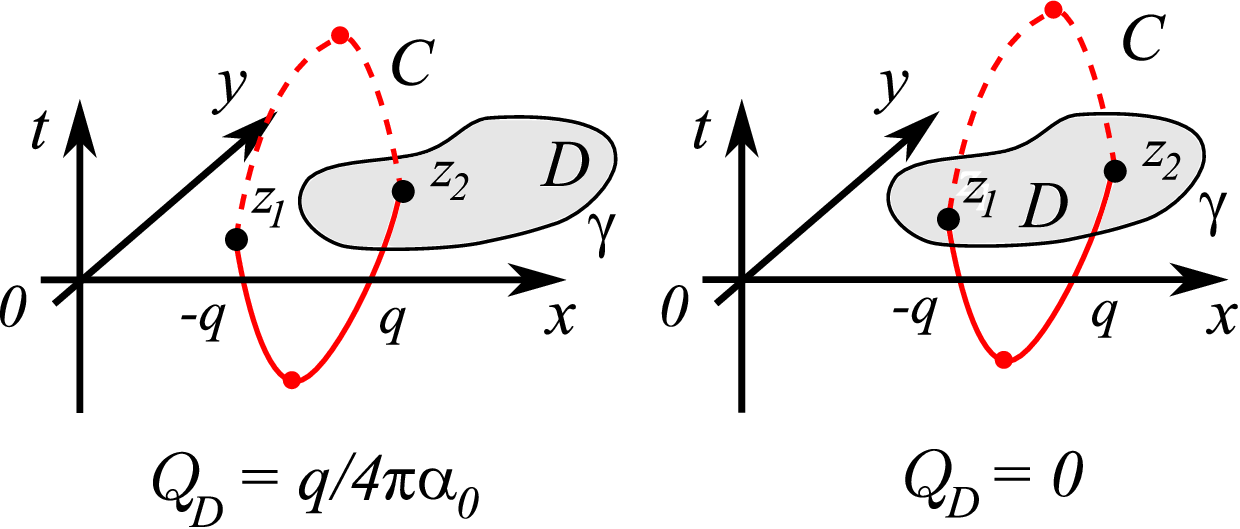}
\caption{Evaluation of the spectrum of the charge operator $Q_D$, defined in Eq.\ (\ref{charge-app}). 
{\it Left panel}: Only one endpoint, $z_2$, is inside region $D$. Eq.\ (\ref{charge-res-app}) then yields $Q_D = q/4\pi\alpha_0$.
{\it Right panel}: Both endpoints $z_1$ and $z_2$ belong to region $D$, and the total charge in $D$ is zero. 
These results are easily understood in terms of the linking number of the boundary $\gamma$ of $D$ 
and the Wilson loop $C$ obtained as a concatenation of the initial (solid) Wilson line with the corresponding time-reversed 
(dashed) line; (see Eq.\ \ref{fcs}).}
\vspace{-3mm}
\label{app-5}
\end{figure}

In order to find the values of charges of the quasi-particles, 
we recall that the connection between the bulk current density of the QH liquid and the Chern-Simons 
field $B_\mu$ is given by Eq.\ (\ref{j-bulk}).
Hence, the charge operator corresponding to a region $D$, namely $Q_D \equiv \int_D d^2z J^0$, can be written as
\begin{equation}
 Q_D = \frac{1}{2\pi}\int_D d^2z\epsilon^{0\mu\nu}\partial_\mu B_\nu = \frac{1}{2\pi}\oint_\gamma B_\mu dx^\mu,
\label{charge-app}
\end{equation}
where we have applied Stokes' theorem, and the contour $\gamma = \partial D$ is the boundary of the region $D$ (see Fig.\ \ref{app-5}).

Next, we formally evaluate the following expectation, using result (\ref{braid-app}): 
\begin{multline}
 \langle\underbar{C}\,|\exp(i\lambda Q_D)|\underbar{C}\,\rangle = \exp\left\{-\frac{\lambda}{4\pi\alpha_0}\sum_i q_i{\cal L}(\gamma, C_i)\right\}\\
 \times \langle\underbar{C}\,|\underbar{C}\,\rangle.
\label{fcs}
\end{multline}
It follows from Eq.\ (\ref{fcs}) and the orthogonality of states with different
end points of Wilson lines that all the sates (\ref{def-states-app}) are
the eigenstates of the charge operator (\ref{charge-app}), and we find that
\begin{equation}
 [Q_D, {\cal W}[\underbar{C}\,]] = -\frac{1}{4\pi\alpha_0}\sum_iq_i\theta(z_i\in D){\cal W}[\underbar{C}\,]\, ,
\label{charge-res-app}
\end{equation}
where the step function $\theta$ is equal to one if the endpoint $z_i$ lies inside the region $D$ and vanishes elsewhere, 
as illustrated in Fig.\ \ref{app-5}.
In other words, each Wilson line innihilates a local excitation of charge $q_i/4\pi\alpha_0$ at the point $z_i$. 
In Sec.\ \ref{bulk-edge}, we have shown that the charges of quasi-particles are quantized. 
This effect resembles the quantization of magnetic flux in superconductors.

\begin{thebibliography}{99}
\bibitem{footnote1}
The use of the term ``charge fractionalization'' in this context is somewhat unfortunate,
because, in contrast to the quasi-particle fractionalization, the corresponding process is completely
classical in nature. In fact, it is very similar to the displacement current in electrical circuits.
\bibitem{footnote2}
Note, in particular, that the average values of physical observables
are not determined solely by the energy spectrum of a system, but also 
by the matrix elements of the observables. Thus, one is not able to make 
a definitive conclusion on the periodicity of observables based on the 
consideration of the spectrum alone.
\bibitem{footnote2-mzf}
We define the statistical phase $\theta_{12}$ of two operators $\psi_1$ and $\psi_2$
via the relation $\psi_1(x)\psi_2(x') = e^{i\theta_{12}}\psi_2(x')\psi_1(x)$. The
microscopic construction \cite{micro} shows that so defined statistical phase indeed
takes integer values for single valued excitations of the Laughlin state. \cite{Laugh}
\bibitem{footnoterho}
Note that the functional integration in Eq.\ (\ref{Z2}) over $\rho(z)$ is
constrained to the domain $\rho(z)\geq 0$.
\bibitem{crystal}
It is expected that 2D electrostatics for an ideal Coulomb plasma
holds for $1/\nu=m< 7$, while for larger ``temperature'' $m$ there is a tendency
to Wigner crystallization. \cite{QHE}
\bibitem{footnote-zeromodes}
To be precise, the operators (\ref{zeromodedef}) are not unitary if $N$ and $M$ are finite.
However, in the thermodynamic limit ($N,M\to\infty$) these operators become unitary and
satisfy the commutation relations (\ref{zeromodecom}).
\bibitem{footnote-com}
This commutation relation follows form the definition of the derivative
operator (\ref{notstandart}),
which differs from the standard one.
\bibitem{footnote-Klein}
We stress, that $\delta M$-dependent phase shifts in 
the individual tunneling operators (\ref{genres}), determined by the 
single-valued functions $\omega_d^+$ and $\omega_d^-$,  are not universal
and cannot be generally represented in terms of the Klein factors introduced
in Ref.\ [\onlinecite{Feldman}].  Only the product 
of two tunneling operators in Eq.\ (\ref{MZ-phase}) acquires 
a universal $\delta M$-dependent phase factor. We show below that this factor 
eventually leads to the suppression of quasi-particle AB oscillations in the 
stationary current through a MZ interferometer.
\bibitem{footnote-ohmic}
On may check that solving the detailed balance equation with tunneling rates calculated
with the tunneling operators (\ref{elec-u}) and (\ref{elec-d}) leads to the distribution
(\ref{rho-eq}).
\bibitem{footnote-app}
In our presentation, we choose readability over mathematical rigor.
A more detailed and rigorous discussion can be found in, e.g., 
Ref.\ [\onlinecite{CS-theory}].
\end{thebibliography}
\end{document}